\documentclass[apj]{emulateapj}
\usepackage{color}
\usepackage{times}
\usepackage{graphicx}
\usepackage{amssymb,amsmath}

\slugcomment{Published by ApJ, April, 2016}

\shorttitle{TTVs near eccentricity-type MMR}
\shortauthors{Deck et al.}

\begin{document}

\title{Transit timing variations for planets near eccentricity-type mean motion resonances}

\author{Katherine M. Deck\altaffilmark{1,3} and Eric Agol\altaffilmark{2,4}}

\altaffiltext{1}{Department of Geological and Planetary Sciences, California Institute of Technology, Pasadena, CA}
\altaffiltext{2}{Department of Astronomy, University of Washington, Seattle, WA}
\altaffiltext{3}{Corresponding author: kdeck@caltech.edu}
\altaffiltext{4}{NASA Astrobiology Institute’s Virtual Planetary Laboratory, Seattle, WA 98195, USA}

\begin{abstract}
We derive the transit timing variations (TTVs) of two planets near a second order mean motion resonance on nearly circular orbits. We show that the TTVs of each planet are given by sinusoids with a frequency of $j n_2-(j-2)n_1$, where $j \ge 3$ is an integer characterizing the resonance and $n_2$ and $n_1$ are the mean motions of the outer and inner planets, respectively. The amplitude of the TTV depends on the mass of the perturbing planet, relative to the mass of the star, and on both the eccentricities and longitudes of pericenter of each planet. The TTVs of the two planets are approximated anti-correlated, with phases of $\phi$ and $\approx \phi+\pi$, where the phase $\phi$ also depends on the eccentricities and longitudes of pericenter. Therefore, the TTVs caused by proximity to a second order mean motion resonance do not in general uniquely determine both planet masses, eccentricities, and pericenters. This is completely analogous to the case of TTVs induced by two planets near a first order mean motion resonance.  We explore how other TTV signals, such as the short-period synodic TTV or a first order resonant TTV, in combination with the second order resonant TTV, can break degeneracies. Lastly, we derive approximate formulae for the TTVs of planets near any order eccentricity-type mean motion resonance; this shows that the same basic sinusoidal TTV structure holds for all eccentricity-type resonances. Our general formula reduces to previously derived results near first order mean motion resonances.
 \end{abstract}
\section{Introduction}
Transit timing variations (TTVs; \citealt{Miralda2002,Agol2005,Holman2005}) have proved useful for constraining the masses and orbital elements of exoplanets (e.g. \citealt{CarterAgol2012,KOI142,Kep56}). To date, most TTV studies have focused on pairs of planets near mean motion resonances (MMRs), and, in particular, on those near first order resonances. This is because near resonances the small perturbations that planets impart on each other can add coherently over time to produce a large, detectable TTV signal, and because for low eccentricity orbits the ``near resonance region" is largest for first order resonances.

{ However, transit timing variation models for planet pairs near resonance are plagued by degeneracies. For a pair of planets on coplanar orbits, there are ten free parameters which control the TTVs. Assuming both planets transit, the periods and initial phases of the orbits are well known, leaving six unknown parameters - the masses, eccentricities, and longitudes of pericenter of the two planets.  \citet{Boue} showed analytically that all of these unknown parameters affect TTV amplitudes and phases for systems near or in eccentricity-type MMRs and concluded that degeneracies between mass and orbital parameters could strongly affect inferences based on TTVs. The particular case of a pair of planets near (but not in) first order resonances was analyzed in detail by \citet{Lithwick2012}, who showed that the TTVs of each planet are approximately sinusoidal, with a period set by the known mean orbital periods.  Furthermore, the first order resonant TTVs are often nearly anti-correlated, in which case there are only three constraining ``observables": two TTV amplitudes and a single phase. The six unknown parameters therefore cannot all be determined uniquely, and in particular a degeneracy between masses and eccentricities results \citep{Lithwick2012}.}

More recently, it has been demonstrated that a small amplitude ``chopping" signal associated with individual planetary conjunctions - and {\it not} with proximity to mean motion resonance - can determine the masses of the interacting planets uniquely, with only weak dependence on the eccentricities and longitudes of pericenter \citep{NesVok,DeckAgol}. If this chopping TTV is measured for a system near a first order resonance, the amplitude and phase of the resonant TTV can be used to constrain the remaining degrees of freedom (the eccentricities $e$ and the longitudes of pericenter $\varpi$). However, as shown by \citet{Lithwick2012}, the individual eccentricities and longitudes of pericenter are not constrained by the first-order resonant TTV; rather, only a linear combination of the eccentricities vectors $(e\cos{\varpi},e\sin{\varpi})$ are (the quantity $Z_{\rm{free}}$ in the notation of \citealt{Lithwick2012}). { This raises the question of if and in which circumstances TTVs can be used to measure individual eccentricities uniquely. }

Here we consider the case of two planets orbiting near a second order resonance, with $P_2/P_1 \approx j$:$j-2$, with $j\ge 3$. Second order resonances distinct from first order commensurabilities appear when $j$ is odd. Though less common than first order MMR in the observed sample of transiting planets \citep{Fabrycky}, these configurations are still of interest and important to understand \citep{Hutter,Petigura}. On the other hand, second order resonances with $j$ even are important because they represent the dominant correction at $O(e^2)$ to the first order resonant TTV formula derived by \citet{Lithwick2012}. Because of the different functional dependence on the eccentricities and longitudes of pericenter between the first and second order terms, second order effects may be important for breaking degeneracies present in the TTVs of planets near first order MMR.  
 
We derive an approximate formula for TTVs resulting from an orbital configuration near a second order resonance in Section \ref{sec:derive}. In Section \ref{sec:degeneracy}, we  interpret the resulting orbital parameter and mass constraints allowed by these TTVs.  In Section \ref{sec:comparison}, we test the TTV formulae in multiple ways. We first compare the predicted TTVs using the formulae with those determined via numerical integration of the full gravitational equations of motion across a wide range of relevant parameter space. Then, for the specific systems Kepler-26 and Kepler-46, we compare the outcome of TTV inversions obtained using the formulae to those obtained via n-body analysis. In Section \ref{sec:fit}, we apply our formula to simulated data and investigate if measuring the second harmonic of the TTVs for a pair of planets near a first order resonance allows unique determinations of both planet eccentricities. For completeness, we extend our derivation to systems near the $j$:$j-N$ $N-$th order mean motion resonance in Section \ref{sec:general}. We give our conclusions in Section \ref{sec:conclude}.

\section{Derivation of the approximate TTV}\label{sec:derive}
We would like to determine approximate expressions for the TTVs induced for two planets near the $j$:$j-2$ second order resonance. To begin, we write the Hamiltonian in Jacobi elements up to second order in planet eccentricities.  We include only the second order resonant terms in this derivation. In this case, the Hamiltonian is
\begin{align}\label{Ham1}
H & = -\frac{GM_\star m_1}{2 a_1}-\frac{GM_\star m_2}{2 a_2}-\frac{G m_1 m_2}{a_2}\times \nonumber \\\bigg[& g_{j,45}(\alpha)e_1^2 \cos{(\theta_j-2\varpi_1)}+g_{j,53}(\alpha)e_2^2 \cos{(\theta_j-2\varpi_2)}\nonumber \\
&+g_{j,49}(\alpha)e_1 e_2 \cos{(\theta_j-\varpi_2-\varpi_1)}\bigg]
\end{align}
where
\begin{align}
\theta_j & = j\lambda_2-(j-2)\lambda_1
\end{align}
and
\begin{align}
g_{j,45}(\alpha) & =\frac{1}{8}((-5j + 4j^2)b_{1/2}^j(\alpha) + (4j - 2)D_\alpha b_{1/2}^j(\alpha)\nonumber \\
&+D_\alpha^2 b_{1/2}^j(\alpha)) ,\nonumber \\
g_{j,49}(\alpha) & =\frac{1}{4}((-2 + 6j - 4j^2) b_{1/2}^{j-1}(\alpha)+ (2- 4j) D_\alpha b_{1/2}^{j-1}(\alpha)  \nonumber \\
&-D_\alpha^2 b_{1/2}^{j-1}(\alpha)), \nonumber \\
g_{j,53}(\alpha) & = \frac{1}{8}((2-7j+4j^2)b_{1/2}^{j-2}(\alpha)+(4j-2)D_\alpha b_{1/2}^{j-2}(\alpha)\nonumber \\
&+D^2_\alpha b_{1/2}^{j-2}(\alpha))-\frac{27\alpha}{8}\delta_{j,3}, \nonumber \\
D^k_\alpha & \equiv \alpha^k \frac{d^k}{d \alpha^k}, \nonumber \\
 b_{1/2}^j(\alpha) & = \frac{1}{\pi} \int_0^{2\pi} \frac{\cos{(j \theta)}}{\sqrt{1-2\alpha \cos{\theta}+\alpha^2}}d \theta.
\end{align}
In the definitions of the $g_{j,xx}$ functions, we have neglected all indirect contributions which arise only at higher orders of eccentricity \citep{MurrayDermott}. Here $a_i$ is the semimajor axis, $e_i$ the orbital eccentricity, $m_i$ the mass, $\lambda_i$ the mean longitude, and $\varpi_i$ the longitude of periastron of the $i-$th planet ($i=1$ or $2$),  $\alpha=a_1/a_2$ and $M_\star$ is the mass of the star. 

{ Throughout this derivation, we will make use of the following small quantities: 
\begin{align}\label{small_params}
\delta &= \frac{\omega_j}{n_i}, \nonumber \\
\epsilon_i &=\frac{m_i}{m_\star},\nonumber \\
e_i&,
\end{align}
where
\begin{align}
n_i & = \frac{G^2 M_\star^2 m_i^3}{ \Lambda_i^3}
\end{align}
and \begin{align}
\omega_j& = jn_2-(j-2)n_1.
\end{align} 
The assumption that the masses are small is required because we neglect terms of order $\epsilon_i^2$ in both Equation \eqref{Ham1} and below, for example in Equation \eqref{F2}- Equation \eqref{homologic}. The assumption of low eccentricities allows us to neglect higher order terms in Equation \eqref{Ham1}, as well as terms at zeroth and linear in eccentricities that are non-resonant. We will further discuss the assumptions of near resonance, $\delta \ll 1$, and of low eccentricities, $e_i\ll 1$, below. For reference, $\delta$ is on the order of a few percent for many systems of interest .}

We will derive the TTVs using an approach developed by \citet{NesMorb}, \citet{N3}, and \citet{NB10} and later used by \citet{DeckAgol}. The method is based on perturbation theory within a Hamiltonian framework, and therefore we need to first convert the orbital elements into canonical variables. Written in terms of the canonical momenta (left) and coordinates (right),
\begin{center}\label{variables}
\begin{tabular}{l c l}
$\Lambda_i  = m_i\sqrt{G M_\star a_i}$ &      &$ \lambda_i$  \\
$x_i  = \sqrt{2P_i}\cos{p_i} $&     &$ y_i =\sqrt{2P_i}\sin{p_i}$ \\
\end{tabular}
\end{center}
with
\begin{center}\label{variables2}
\begin{tabular}{lcr}
$P_i = \Lambda_i \frac{e_i^2}{2} + O(e_i^4) $&     & $p_i = -\varpi_i,$
\end{tabular}
\end{center}
Equation \eqref{Ham1} can be rewritten as
\begin{align}\label{Ham2}
H & =H_0(\Lambda_1,\Lambda_2)+ H_1, \nonumber \\
H_0 & =  -\frac{G^2 M_\star^2 m_1^3}{2 \Lambda_1^2}-\frac{G^2 M_\star^2 m_2^3}{2 \Lambda_2^2},\nonumber \\
H_1 & =-\epsilon_1\frac{G^2 M_\star^2 m_2^3}{ \Lambda_2^2}\bigg[\tilde{A}_1 \cos{\theta_j} + \tilde{A}_2 \sin{\theta_j}\bigg], \nonumber \\
\tilde{A}_1 & = \frac{g_{j,45}}{\Lambda_1}(x_1^2-y_1^2) + \frac{g_{j,53}}{\Lambda_2}(x_2^2-y_2^2)+ \frac{g_{j,49}}{\sqrt{\Lambda_1 \Lambda_2}}(x_1x_2-y_1y_2), \nonumber \\
\tilde{A}_2 & = -\frac{g_{j,45}}{\Lambda_1}2x_1y_1 -\frac{g_{j,53}}{\Lambda_2}2 x_2 y_2- \frac{g_{j,49}}{\sqrt{\Lambda_1 \Lambda_2}}(x_1y_2+x_2y_1).
\end{align}
The coefficients $g_{j,xx}$ are evaluated at semimajor axis ratio $\alpha(\Lambda_1,\Lambda_2)= (\Lambda_1/m_1)^2(m_2/\Lambda_2)^2$.

Our goal is to find a new set of canonical variables (denoted with primes) such that in the new set the Hamiltonian takes the form
\begin{align}\label{Ham2_new}
H' & =H_0(\Lambda_1',\Lambda_2')=  -\frac{G^2 M_\star^2 m_1^3}{2 \Lambda_1^{'2}}-\frac{G^2 M_\star^2 m_2^3}{2 \Lambda_2^{'2}}.
\end{align}

In the new variables, the motion of the two planets is ``Keplerian". These Keplerian orbits correspond to the average of the perturbed orbits (averaged over the periodic terms in $H_1$). Transit timing variations are deviations from the times predicted from the mean ephemeris of a planet. In practice, we estimate this by fitting the transit times with a constant period (Keplerian) model. The deviations are caused by the interaction with the other planet, and hence the transformation we seek to turn Equation \eqref{Ham1} into Equation \eqref{Ham2} is precisely what we need to give us the TTVs.

To determine this transformation, we use a Type-2 generating function of the form:
\begin{align}\label{F2}
F_2(\lambda_i,y_i,\Lambda'_i,x_i') &= \lambda_i \Lambda'_i + y_i x'_i + \it{f}(\lambda_i,y_i,{\Lambda}'_i,x'_i),
\end{align}
which relates the old and new variables as
\begin{align}\label{transform}
\Lambda_i & = \frac{\partial F_2}{\partial \lambda_i} = \Lambda_i'+\frac{\partial \it{f}}{\partial \lambda_i}, \nonumber \\
x_i & = \frac{\partial F_2}{\partial y_i} = x_i'+\frac{\partial \it{f}}{\partial y_i}, \nonumber \\
\lambda_i' & = \frac{\partial F_2}{\partial \Lambda_i'} = \lambda_i+\frac{\partial \it{f}}{\partial \Lambda_i'}, \nonumber \\
y_i' & = \frac{\partial F_2}{\partial x_i'} = y_i+\frac{\partial \it{f}}{\partial x_i'} .
\end{align}
Therefore the first piece of $F_2$ is the identity transformation, and the second piece $f$  is a small correction which will be linear in $\epsilon_1$. The function $f$ which produces the new ``Keplerian" Hamiltonian of Equation \eqref{Ham2_new} from the Hamiltonian of Equation \eqref{Ham2} is 
\begin{align}\label{ffunc}
f &= \epsilon_1\frac{n_2 \Lambda_2}{\omega_j}\bigg[\tilde{A}_1 \sin{\theta_j} - \tilde{A}_2 \cos{\theta_j}\bigg],
\end{align}

which we determine by solving the homologic equation
\begin{align}\label{homologic}
\bigg\{ f,H_0\bigg\}+H_1 =0
\end{align}
where $\{\ldots,\ldots\}$ denotes a Poisson Bracket. For more details on this type of derivation, one can refer to \citealt{MorbidelliBook}, \citealt{NesMorb} or \citealt{DeckAgol}.  Formally, $f$ is a mixed-variable function of both old coordinates and new momenta. In Equation \eqref{ffunc}, however, we neglect to make this distinction. This is because the difference between the two sets, given implicitly in Equation \eqref{transform}, depends on derivatives of $f$, which is itself linear in $\epsilon_1$. Therefore, within $f$ itself, the difference between the two sets is negligible since we are only working to first order in $\epsilon_1$.

Equation \eqref{transform} shows us how to derive, using the function $f$ of Equation \eqref{ffunc}, the difference between the real and averaged canonical variables. The results are
\begin{align}\label{deviations}
\lambda_i -\lambda_i' & \equiv \delta \lambda_i \approx \epsilon_1\frac{n_2 \Lambda_2}{\omega_j^2}\frac{d\omega_j}{d\Lambda_i}\bigg[\tilde{A}_1 \sin{\theta_j} - \tilde{A}_2 \cos{\theta_j}\bigg],  \nonumber \\
y_i -y_i' & \equiv \delta y_i = - \epsilon_1\frac{n_2 \Lambda_2}{\omega_j}\bigg[\frac{d\tilde{A}_1}{d x_i} \sin{\theta_j} - \frac{d\tilde{A}_2}{d x_i} \cos{\theta_j}\bigg],\nonumber \\
x_i -x_i' & \equiv \delta x_i  =\epsilon_1\frac{n_2 \Lambda_2}{\omega_j}\bigg[\frac{d\tilde{A}_1}{d y_i} \sin{\theta_j} - \frac{d\tilde{A}_2}{d y_i} \cos{\theta_j}\bigg], \nonumber\\
\Lambda_i -\Lambda_i' & \equiv \delta \Lambda_i  \approx 0.
\end{align}
{ Here we have made use of the small parameters given in Equation \eqref{small_params}. That is, we have assumed that since the pair is close to resonance with small eccentricities, we need only retain terms of order $e/\delta$ and $(e/\delta)^2$. Terms proportional to $e^2/\delta$ or missing the small denominator $\delta$ are assumed to be considerably smaller and can be neglected. This approximation holds as long as $e \ll 1$ and as long as $\delta \ll 1$. The dominant neglected term $e^2/\delta$ is negligible compared with the synodic chopping terms (of zeroth order in $e$ and $1/\delta$) for $e \lesssim \sqrt{\delta}$. For $\delta$ of a few percent, as for many of the {\it Kepler} systems, this corresponds to $e \lesssim 0.1$.}

We now must take changes in the canonical elements and convert them into TTVs.  The transit occurs when the true anomaly of the planet $\theta$ is equal to a value, which, given the reference frame, aligns the planet in front of the star along our line of sight. $\theta$ is not a canonical variable, but it can be related to our canonical set via a power series in eccentricity of the transiting planet:
\begin{align}\label{Eqn2}
\theta[\lambda,\Lambda,x,y] & = \lambda + \frac{2}{\sqrt{\Lambda}}\bigg( x \sin{\lambda}+y \cos{\lambda} \bigg)+O(e^2)+\ldots
\end{align}
Perturbing Equation \eqref{Eqn2} about the averaged orbit yields
\begin{align}\label{Eqn3a}
\delta \theta & = \frac{\partial\theta[\lambda,\Lambda,x,y]}{\partial\lambda}\delta \lambda+  \frac{\partial\theta[\lambda,\Lambda,x,y]}{\partial x}\delta x +  \frac{\partial \theta[\lambda,\Lambda,x,y]}{\partial y}\delta y
\end{align}
where we have already neglected the $\delta \Lambda$ piece with the knowledge that it will be small. The derivatives of $\theta[\lambda,\Lambda,x,y]$ with respect to any of the remaining variables will be proportional to $e^0, e^1, e^2$ - without any small denominators. Hence we also only keep $\partial \theta[\lambda,\Lambda,x,y]/\partial x$, $\partial \theta[\lambda,\Lambda,x,y]/\partial y$, and $\partial \theta[\lambda,\Lambda,x,y]/\partial \lambda $ to zeroth order in $e$ since we have assumed eccentricities are small. We approximate
\begin{align}\label{Eqn3}
\delta \theta & \approx \delta \lambda+ \frac{2}{\sqrt{\Lambda}}\bigg( \delta x \sin{\lambda}+\delta y \cos{\lambda}  \bigg) +\ldots
\end{align}

To turn Equation \eqref{Eqn3} into a timing perturbation, we need to convert  $\delta \theta$ into $\delta t$. This is achieved by relating $\theta$ to $\lambda$, since $\lambda$ is a linear function of time. We can write
\begin{align}\label{transformAgain}
\delta \theta & = -n \delta t +O(e)+\ldots,
\end{align}
where {\it again} we can neglect the $O(e)$ correction, since this is a factor of $e$ without a small denominator $\delta = \omega_j/n_i$. Combining the results of Equations \eqref{deviations}, Equation \eqref{Eqn3} and Equation \eqref{transformAgain}, we find that the TTVs are approximately given by:
\begin{align}\label{ttvs_a_b}
\delta t_1&  = -\frac{1}{n_1}\epsilon_2 \alpha\bigg[\bigg\{3(j-2)\bigg(\frac{n_1}{\omega_j}\bigg)^2 A_1-2\bigg(\frac{n_1}{\omega_j}\bigg)B^1_1\bigg\}\sin{\theta_j}\nonumber \\
&+\bigg\{3(j-2)\bigg(\frac{n_1}{\omega_j}\bigg)^2 A_2-2\bigg(\frac{n_1}{\omega_j}\bigg)B^1_2\bigg\}\cos{\theta_j}\bigg]
\end{align}
and
\begin{align}\label{ttvs_a_b2}
\delta t_2  &= -\frac{1}{n_2}\epsilon_1 \bigg[\bigg\{-3j\bigg(\frac{n_2}{\omega_j}\bigg)^2 A_1-2\bigg(\frac{n_2}{\omega_j}\bigg)B^2_1\bigg\}\sin{\theta_j}\nonumber \\
&+\bigg\{-3j\bigg(\frac{n_2}{\omega_j}\bigg)^2 A_2-2\bigg(\frac{n_2}{\omega_j}\bigg)B^2_2\bigg\}\cos{\theta_j}\bigg]
\end{align}
where
\begin{align}
A_1 & = \tilde{A}_1 = g_{j,45}(\alpha)e_1^2 \cos{(2\varpi_1)}+ g_{j,53}(\alpha)e_2^2 \cos{(2\varpi_2)}\nonumber \\
&+ g_{j,49}(\alpha)e_1 e_2 \cos{(\varpi_1+\varpi_2)} \nonumber \\
A_2 & = - \tilde{A}_2 = -g_{j,45}(\alpha)e_1^2 \sin{(2\varpi_1)}-g_{j,53}(\alpha)e_2^2 \sin{(2\varpi_2)}\nonumber \\
&- g_{j,49}(\alpha)e_1 e_2 \sin{(\varpi_1+\varpi_2)}\nonumber \\
B_1^1 & = 2 g_{j,45}(\alpha) e_1\cos{(\lambda_1+\varpi_1)}+ g_{j,49}(\alpha) e_2\cos{(\lambda_1+\varpi_2)}\nonumber \\
B_2^1 & = -2 g_{j,45}(\alpha) e_1\sin{(\lambda_1+\varpi_1)}-g_{j,49}(\alpha) e_2\sin{(\lambda_1+\varpi_2)}\nonumber \\
B_1^2 & = 2 g_{j,53}(\alpha) e_2\cos{(\lambda_2+\varpi_2)}+ g_{j,49}(\alpha) e_1\cos{(\lambda_2+\varpi_1)}\nonumber \\
B_2^2 & = -2 g_{j,53}(\alpha) e_2\sin{(\lambda_2+\varpi_2)}-g_{j,49}(\alpha) e_1\sin{(\lambda_2+\varpi_1)}.
\end{align}

{ When will these formulae break down? Here we give some qualitative expectations, before turning to numerical tests in Section \ref{sec:comparison}. 

Although we are considering the near resonance case, if $\delta$ become too small neglected terms proportional to $1/\delta$ will become important if the system is too close to resonance. Additionally, our formulae will not apply when a system is in resonance. The width of the resonance grows with $e$ and with $\epsilon$. For a given $\epsilon$, then, we expect our formulae to fail if the eccentricity is too large, either because of being in resonance or because of neglected higher order terms.

As $\alpha \rightarrow 1$, the functions of Laplace coefficients appearing as coefficients in the disturbing function can diverge, mitigating the effect of small eccentricities raised to high powers. In practice, this means that the derived formulae incur larger error as $\alpha \rightarrow 1$ due to these neglected higher order in $e$ terms. 

Throughout this derivation, we neglected terms without a small denominator $\delta$. In the low eccentricity regime, the dominant contribution comes from the terms independent of eccentricity.  If $e\lesssim \delta$, these zeroth order (in $e$) chopping terms will be comparable in magnitude to the second order TTV. In practice, this implies that for many systems the second order resonant formula should be used with the chopping formula of \citet{NesVok} or \citet{DeckAgol}. Of course, if the second order MMR under consideration is a $O(e^2)$ correction of a first order MMR, one must include the first order resonant contributions (not presented here) as well. 

Note that these transit timing variations were derived in terms of Jacobi elements, which, for the outer planet, are not the defined relative to the star but to the center of mass of the inner-planet-star system. However, true transits occur with respect to the star, not the center of mass of the inner subsystem. The necessary correction be determined by treating the motion of the star as a sum of two Keplerian orbits (e.g. \citealt{Agol2005}). However, the indirect contributions resulting from this correction do not have small denominators, and hence they are not important at this level of approximation. 
} 


\section{Interpretation of the approximate TTV}\label{sec:degeneracy}
The TTV expressions given in Equation \eqref{ttvs_a_b} and Equation \eqref{ttvs_a_b2} depend on the eccentricities and longitudes of pericenter of each planet, after averaging over the TTV period and the orbital periods of the planets\footnote{To be clear, the ``average" eccentricity referred to here and below is the eccentricity computed from the average canonical variables $x$ and $y$, which differs in from the average in time of the eccentricity.}. The only further variation in these quantities is due to secular evolution.  If we assume that the observational baseline is short compared to the secular timescale, they will be approximately constant.  We now show that the TTVs approximately depend only on the masses of the two planets (relative to the mass of the host star) and the approximately constant quantities 
\begin{align}
\delta k & = k_1-k_2\nonumber \\
\delta h & = h_1-h_2
\end{align}
where $h_i = e_i\sin{\varpi_i}$ and $k_i = e_i\cos{\varpi_i}$. 

We define 
\begin{align}
\Delta_j&\equiv \frac{j-2}{j}\frac{P_2}{P_1}-1 = -\frac{\omega_j}{j n_2},
\end{align}
and, substituting in, the TTVs become
\begin{align}\label{ttv1}
\delta t_1  = -\frac{2}{n_1\alpha^{1/2} j \Delta_j}\epsilon_2 \bigg[&\bigg\{\frac{3(j-2)}{2j \Delta_j}\alpha^{-3/2} A_1+B^1_1\bigg\}\sin{\theta_j}\nonumber \\
&+\bigg\{\frac{3(j-2)}{2j \Delta_j}\alpha^{-3/2}  A_2+B^1_2\bigg\}\cos{\theta_j}\bigg]
\end{align}
and
\begin{align}\label{ttv2}
\delta t_2  = -\frac{2}{j n_2\Delta_j}\epsilon_1 \bigg[&\bigg\{-\frac{3}{2\Delta_j} A_1+B^2_1\bigg\}\sin{\theta_j}\nonumber \\
&+\bigg\{-\frac{3}{2\Delta_j} A_2+B^2_2\bigg\}\cos{\theta_j}\bigg]
\end{align}
\begin{figure}[h!]
	\begin{center}
	\includegraphics[width=\columnwidth]{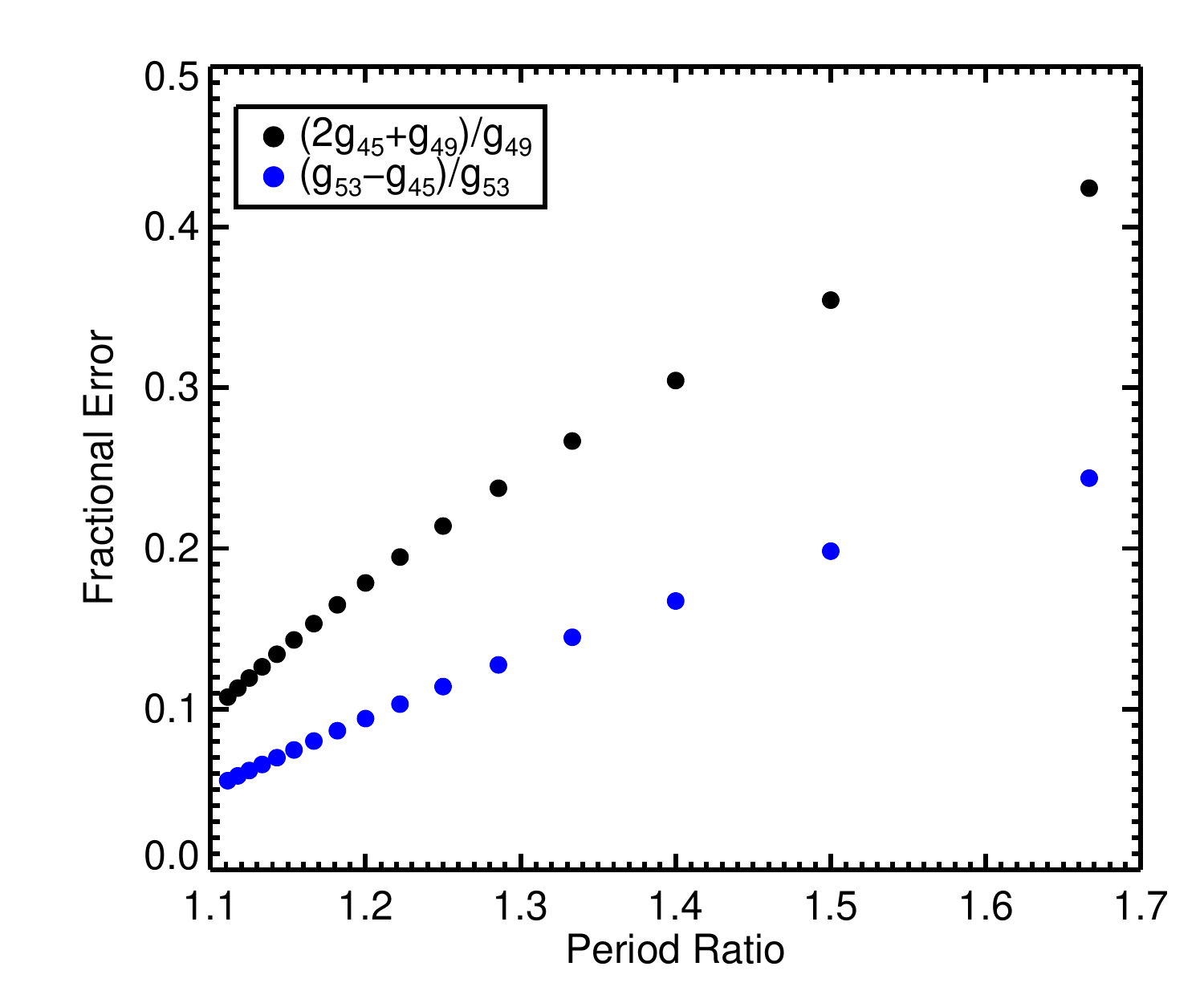}
	\caption{Validity of the approximation that  $g_{j,53}(\alpha) \approx g_{j,45}(\alpha)$ and that $g_{j,49}(\alpha) \approx -2 g_{j,45}(\alpha) $}
	\label{fig1}
	\end{center}
	\end{figure}
Next, we choose our reference frame such that the true longitude $\theta$ at transit is zero\footnote{Note that we are working in Jacobi coordinates, and that in reality the transit occurs when the true longitude in astrocentric coordinates is zero. However, here we make an approximation to $\lambda$ at transit at the TTV level, which is already of order $\epsilon$. The correction from Jacobi to astrocentric coordinates for $\lambda$ at transit, of order $\epsilon$, therefore produces an $\epsilon^2$ correction which we can safely ignore.}.  Then, in accordance with our previous neglect of terms of $O(e)$ without a small denominator $\delta$, the mean longitude at transit is also zero.


Now, as shown by Figure \ref{fig1}, with an error of only a factor of 1-2 one can approximate  $g_{j,53}(\alpha) \approx g_{j,45}(\alpha)$ and $g_{j,49}(\alpha) \approx -2 g_{j,45}(\alpha)$ (shown evaluated at $\alpha = [(j-2)/j]^{2/3}$), in which case
\begin{align}
A_1 &\approx g_{j,45}(\alpha)[\delta k^2-\delta h^2] \nonumber \\
A_2 &\approx -2g_{j,45}(\alpha)\delta k\delta h\nonumber \\
B_1^1 & \approx 2g_{j,45}(\alpha)\delta k \nonumber \\
B_2^1 & \approx -2g_{j,45}(\alpha)\delta h \nonumber \\
B_1^2 & \approx -2g_{j,45}(\alpha)\delta k= -B_1^1 \nonumber \\
B_2^2 &\approx2g_{j,45}(\alpha)\delta h =-B_2^1.
\end{align}

Then the TTVs are roughly given by
\begin{align}
\delta t_1  &= -\frac{2g_{j,45}}{n_1\alpha^{1/2} j \Delta_j}\epsilon_2 \bigg[\bigg\{\frac{3}{2 \Delta_j} [\delta k^2-\delta h^2]+2\delta k\bigg\}\sin{\theta_j}\nonumber \\
&-\bigg\{\frac{3}{\Delta_j}  \delta k\delta h+2\delta h\bigg\}\cos{\theta_j}\bigg]
\end{align}
and
\begin{align}
\delta t_2  &= -\frac{2g_{j,45}}{j n_2\Delta_j}\epsilon_1 \bigg[-\bigg\{\frac{3}{2\Delta_j} [\delta k^2-\delta h^2]+2\delta k\bigg\}\sin{\theta_j}\nonumber \\
&+\bigg\{\frac{3}{\Delta_j} \delta k\delta h+2\delta h\bigg\}\cos{\theta_j}\bigg],
\end{align}
where we have also approximated $\alpha^{-3/2} \approx j/(j-2)$.
If we set
\begin{align}
\cos{\phi}& = \frac{\bigg\{\frac{3}{2 \Delta_j} [\delta k^2-\delta h^2]+2\delta k\bigg\}}{\mathcal{N}},\nonumber \\
\sin{\phi}& = \frac{\bigg\{\frac{3}{\Delta_j}  \delta k\delta h+2\delta h\bigg\}}{\mathcal{N}},\nonumber
\end{align}
and
\begin{align}
\mathcal{N}^2 & = \bigg\{\frac{3}{2 \Delta_j} [\delta k^2-\delta h^2]+2\delta k\bigg\}^2+\bigg\{\frac{3}{\Delta_j}  \delta k\delta h+2\delta h\bigg\}^2,
\end{align}
then
\begin{equation}
\delta t_1  = -\frac{P_1g_{j,45}\alpha^{-1/2}}{j\pi \Delta_j}\epsilon_2 \mathcal{N} \sin{(\theta_j-\phi)}
\end{equation}
and
\begin{equation}
\delta t_2  = \frac{P_2 g_{j,45}}{j \pi \Delta_j}\epsilon_1\mathcal{N} \sin{(\theta_j-\phi)}.
\end{equation}
Therefore, the TTVs of a pair of planets near a second order resonance with low eccentricity are approximately given by sinusoidal motion with a phase $\phi$ set by the eccentricities and pericenters in the combinations of $\delta k =e_1\cos{\varpi_1}-e_2\cos{\varpi_2}$ and $\delta h=e_1\sin{\varpi_1}-e_2\sin{\varpi_2}$. The amplitude is determined by both the mass of the perturbing planet and a factor depending again on $\delta k$ and $\delta h$. Lastly, the TTVs of the two planets are anti-correlated. Assuming both planets transit ($\alpha,\Delta_j, P_1,P_2$ and the time evolution of $\theta_j$ are known), the only unknowns are $\epsilon_1,\epsilon_2,\mathcal{N}$, and $\phi$. However, from the TTVs alone, we only obtain two amplitudes and a phase. 

This outcome is similar to that of the TTVs of a pair of planets near first order resonances, where only the combinations $\delta h $ and $\delta k$ appeared in the TTVs (through the real and imaginary parts of $Z_{free}=f e_1 e^{i\varpi_1}+g e_2 e^{i\varpi_2}$, with $f\approx -g$). The relation $f\approx -g$ is analogous to the approximation we made here regarding Laplace coefficients and the combinations $g_{j,45}\approx g_{j,53}$ and $g_{j,49}\approx -2g_{j,45}$. Again in that case the observables are two amplitudes and a single phase because the TTVs are approximately anti-correlated.

 One way to break degeneracies is to measure other independent harmonics in the TTVs. For example, one might measure the ``chopping" signal and determine $\epsilon_1$ and/or $\epsilon_2$ from that. In reality, there will be a nonzero phase offset, and if it is measured significantly, the TTVs yield four observables, assuming the mean ephemerides are known because each planet transits. However, this is still not enough to determine both eccentricities and longitudes of pericenter uniquely.  It could be that chopping effects which appear at first order in the eccentricities allows individual eccentricities  and longitudes to be measured, though these are small amplitude.

For a pair of planets near a first order resonance, the TTV derived above will represent an O($e^2$) correction to the TTVs. For example, a pair near the $k$:$k-1$ resonance will exhibit TTVs with a period equal to $2\pi/|kn_2-(k-1)n_1|$ and also at the second harmonic, with a period of $2\pi/|jn_2-(j-2)n_1|$ (with $j=2k$). This second harmonic appears with a different dependence on $\delta h$ and $\delta k$, and if measured, could allow for unique mass measurements as well. However, since both the first and second order harmonics depend approximately on $\delta h$ and $ \delta k$, the higher order eccentricity corrections to the formula of \citep{Lithwick2012} may not allow for individual eccentricity measurements, especially for low signal to noise data. However, if the relative phase offsets of either harmonic from $\pi$ can be measured, this second order harmonic could in theory allow for unique measurements of both eccentricities and pericenters as well. We test this in Section \ref{sec:fit}.

\section{Numerical tests of the formula}\label{sec:comparison}

\subsection{Comparisons with direct n-body integration}

We have carried out a comparison of the second-order formula with
N-body simulations of TTVs carried out with {\it TTVFast} \citep{TTVFast}.
We simulated a system of two planets with $m_1/m_\star = m_2/m_\star = 10^{-5}$
with aligned longitudes of pericenter, $\varpi_1 = \varpi_2$, and anti-aligned longitudes of pericenter, $\varpi_1=\varpi_2 + \pi$. The initial phases and $\varpi_1$ were chosen randomly. 
The TTVs determined from the n-body simulation were computed for an inner planet
period of 30 days, over a duration of 1600 days, to mimic a typical transiting
planet system in the Kepler dataset. The
eccentricity vectors were held fixed at the value computed from the
N-body simulation averaged over 1600 days.  The ephemerides were allowed to
vary, and were varied to optimize the agreement between the n-body and
analytic formula, while $\alpha$ used in computing the coefficients was
given by $\alpha = (P_1/P_2)^{2/3}$, where $P_1$ and $P_2$ are the periods
fit to give the ephemerides.

\begin{figure}
\centering
\includegraphics[width=\columnwidth]{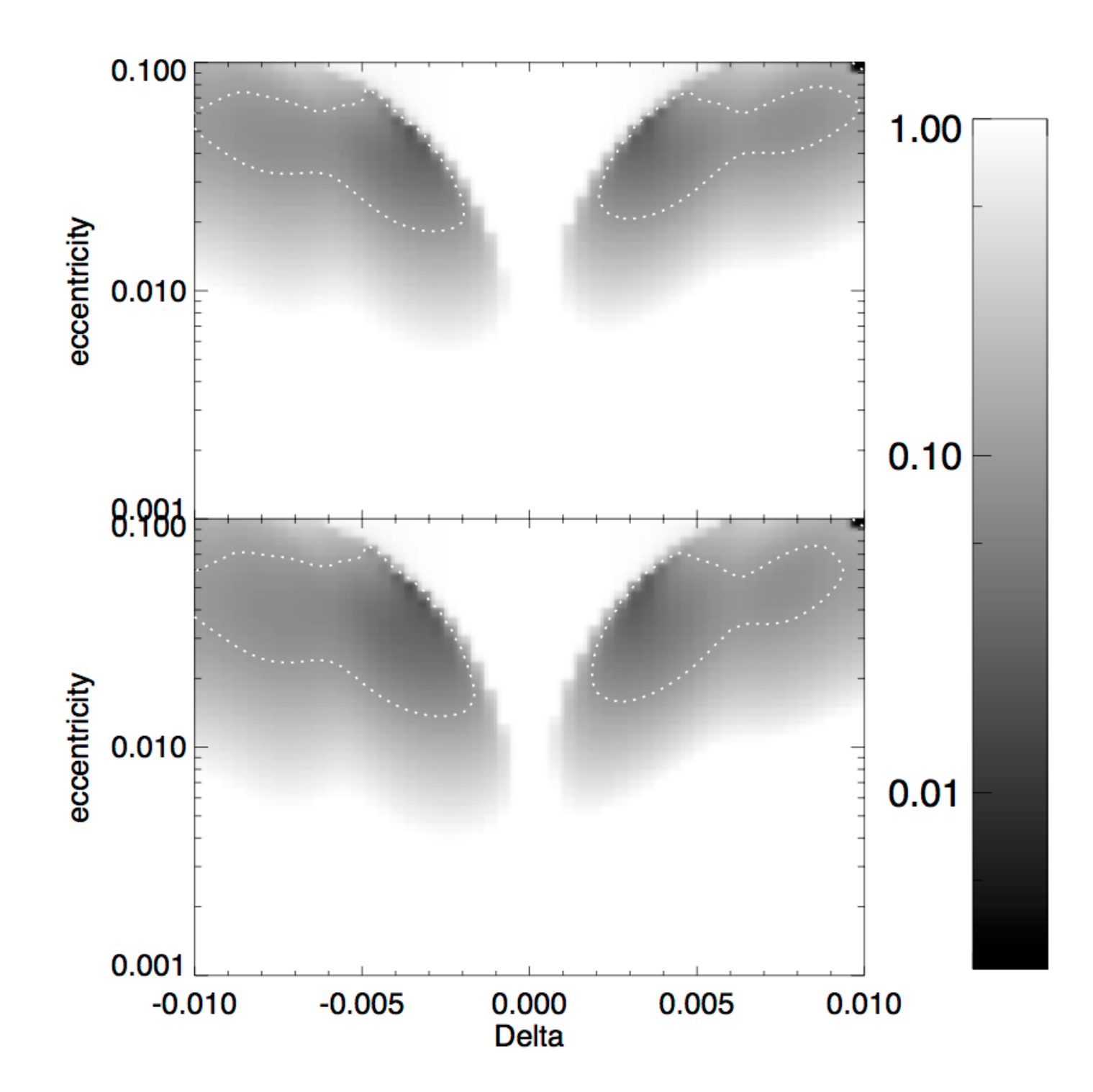}
\caption{Comparison of {\it TTVFast} with the second-order
formula, near the 5:3 second order resonance. Error is given by the standard deviation of the residuals of
the analytic fit to the numerical TTVs, divided by the standard deviation
of the TTVs. The top panel shows the result for the inner planet, the bottom is for the outer planet. 
Dotted lines: 10\% error. 
}
\label{fig:53only}
\end{figure}

We first focused on a range of $\Delta(\alpha)$ near the 5:3 second order resonance. In Figure \ref{fig:53only}, we show, for anti-aligned pericenters, the fractional error in the formula given in Equation \eqref{ttvs_a_b} and in Equation \eqref{ttvs_a_b2}. The error is less than 10\% only for a small range in eccentricity, but importantly the region where the second order formula alone applies is qualitatively as expected. For low eccentricities, the chopping terms are as large as the second order resonant terms, and they are neglected in this fit (as discussed at the end of Section \ref{sec:derive}). For larger eccentricities, this narrow range of $\Delta$ includes resonant orbits, which our formulae does not apply for. 

\begin{figure}
\centering
\includegraphics[width=\columnwidth]{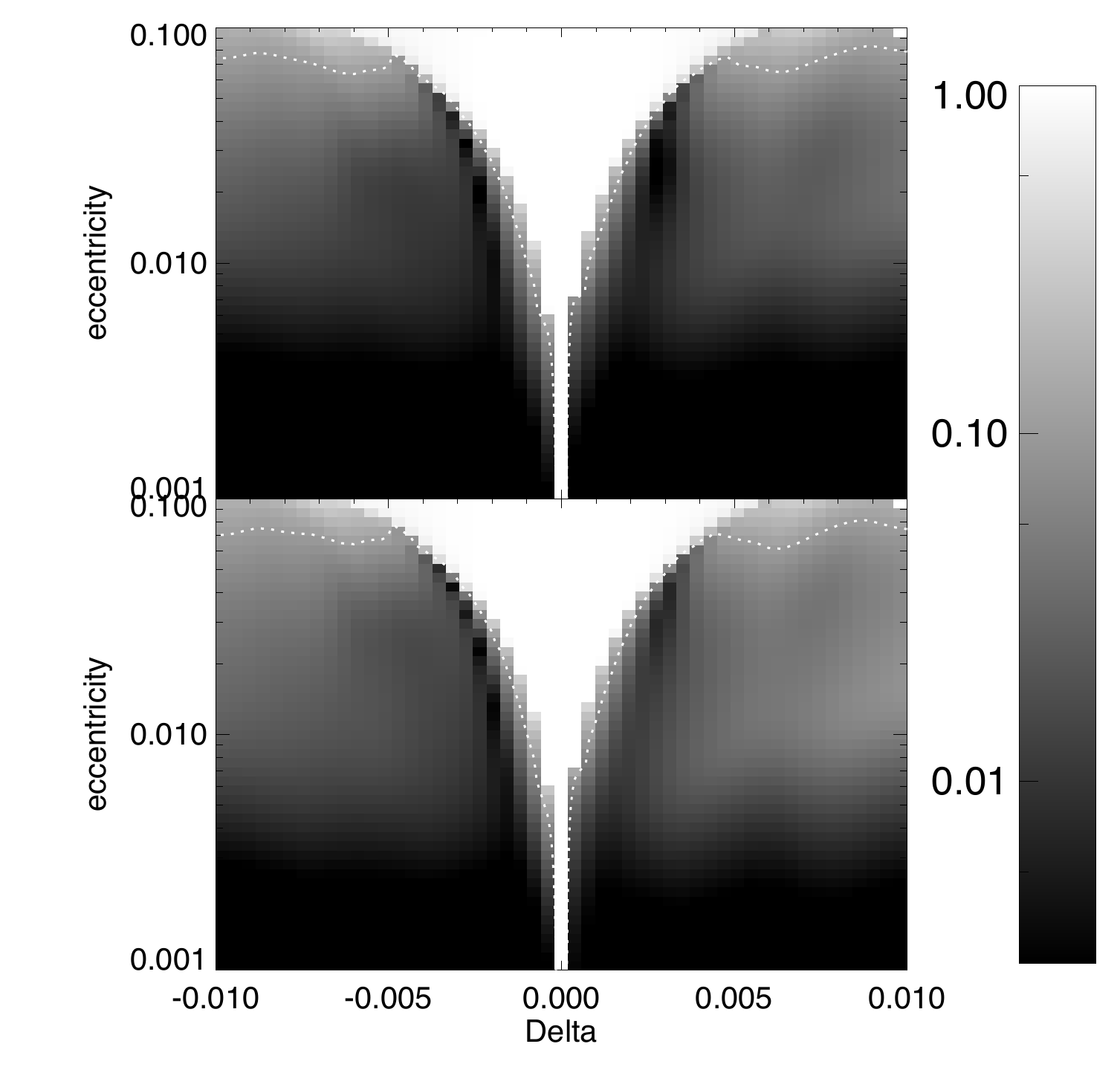}
\caption{Comparison of {\it TTVFast} with the first-order plus second-order
formula, near the 5:3 second order resonance. Error is given by the standard deviation of the residuals of
the analytic fit to the numerical TTVs, divided by the standard deviation
of the TTVs. The dotted line again indicates the 10\% error level.}
\label{fig:zoom5to3}
\end{figure}

We also tried fitting the numerically determined TTVs with the second-order formula added to the first-order formulae presented in \citealt{Agol2015}. Note that the first order formulae include {\it all} terms linear in the eccentricity, while the second order formulae only include the near resonant pieces.  In Figure \ref{fig:zoom5to3}, we show the resulting comparison between the extended formula and the n-body results. The agreement is now excellent even at low eccentricity, as expected since we have included the chopping terms. However, there is still a clear resonant region where our formulae fails.

\begin{figure}
\centering
\includegraphics[width=\columnwidth]{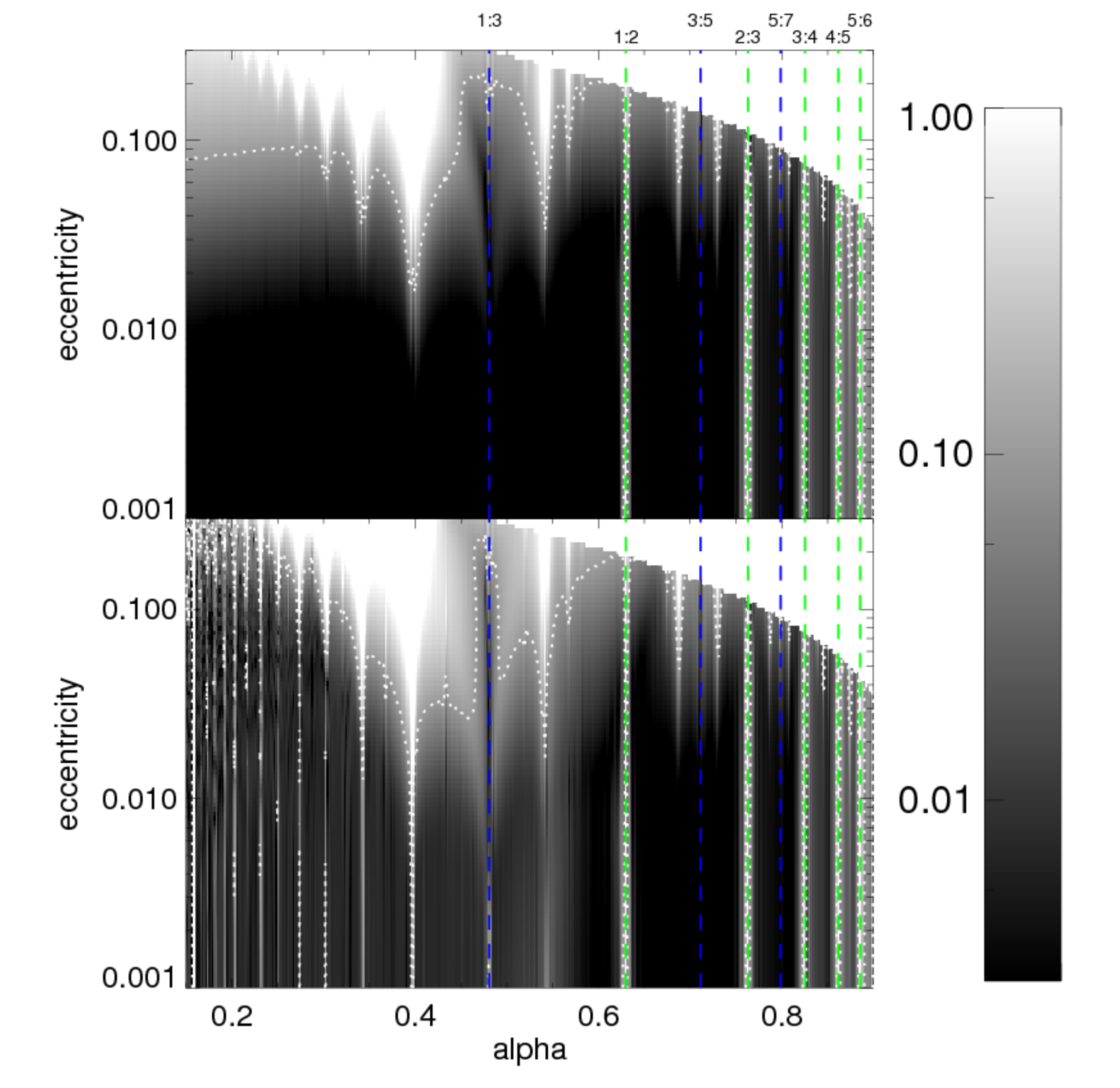}
\caption{Comparison of {\it TTVFast} with the first-order plus second-order
formula. Error is given by the standard deviation of the residuals of
the analytic fit to the numerical TTVs, divided by the standard deviation
of the TTVs, with the longitudes of periastron aligned ($\varpi_1=\varpi_2$). The top panel shows the result for the inner planet, the bottom is for the outer planet. 
Dotted lines: 10\% error level.  Upper right:
 Hill unstable models were not computed, and show 100\%
error.  Green dashed lines: locations
of $j$:$j-1$ resonances; blue dashed lines: $j$:$j-2$ resonances.}
\label{fig:second_order_comparison_aligned}
\end{figure}

\begin{figure}
\centering
\includegraphics[width=\columnwidth]{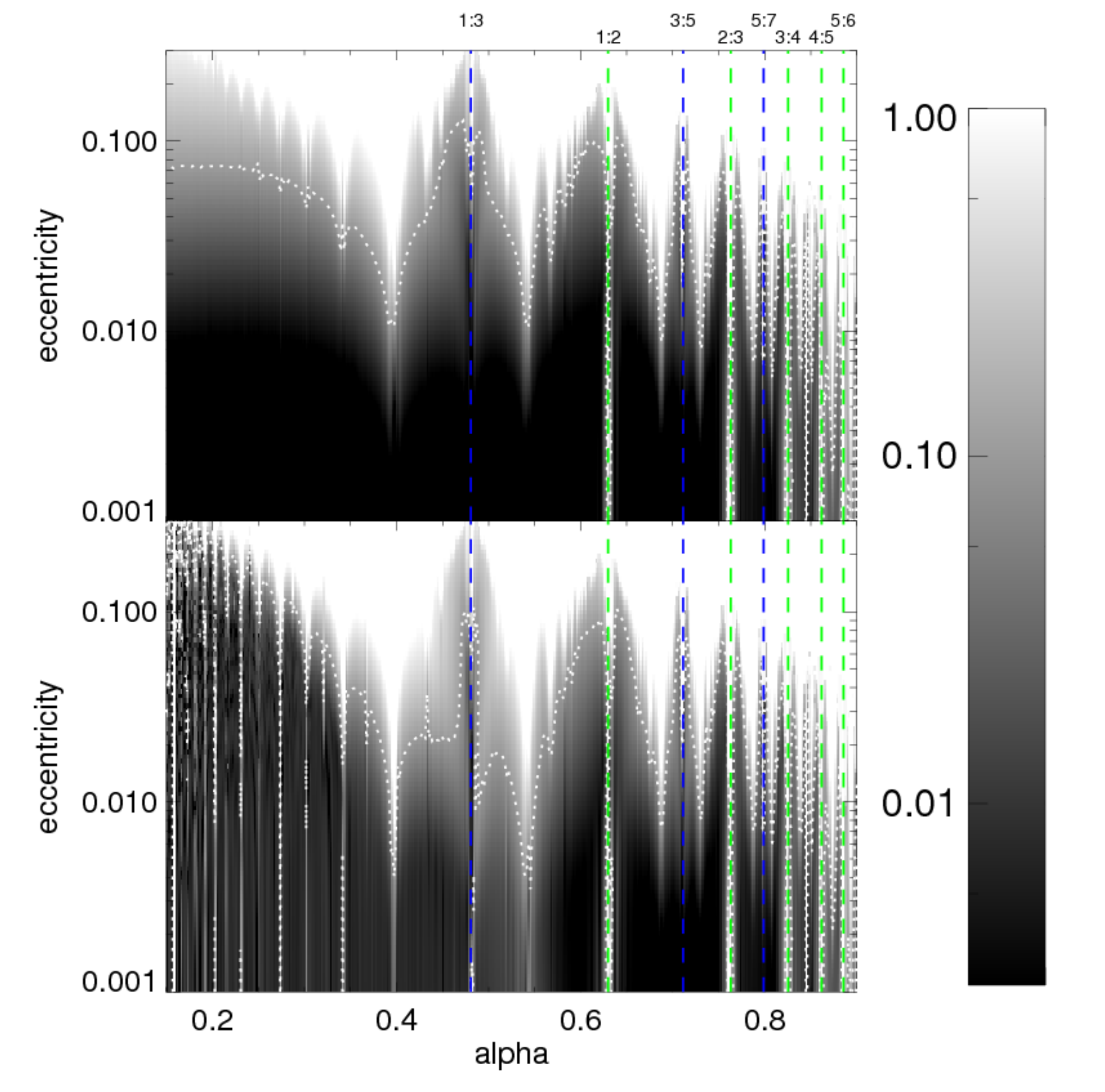}
\caption{Comparison of {\it TTVFast} with the first-order plus second-order
formula.  Error is given by the standard deviation of the residuals of
the analytic fit to the numerical TTVs, divided by the standard deviation
of the TTVs, with the longitudes of periastron anti-aligned ($\varpi_1=\varpi_2+\pi$).
The labels and lines are the same as in Figure \ref{fig:second_order_comparison_aligned}.}
\label{fig:second_order_comparison_antialigned}
\end{figure}

Figures \ref{fig:second_order_comparison_aligned} and 
\ref{fig:second_order_comparison_antialigned} show the results of the aligned
and anti-aligned longitudes of periastron, now for a much larger range of $\alpha$.  The anti-aligned longitudes 
of periastron tends
to maximize the discrepancy, while the aligned tends to minimize.
The fractional precision of the
model was computed from the scatter of the residuals of the fit divided into
the scatter in the n-body TTVs. The mean longitudes and the longitude of
periastron of the inner planet were chosen randomly, and do not affect the 
appearance of this plot significantly.

We found that including the second-order term improves the fit to the n-body
simulation significantly near $j$:$j-2$ period ratios (as demonstrated also by Figure \ref{fig:53only} and Figure \ref{fig:zoom5to3}), which are indicated
in the plot, allowing the analytic formulae to be used to much higher eccentricity
than in the case of the first-order formula only \citep{Agol2015}.  It also improves the
fit near the $j$:$j-1$ resonances as each of these is close to a $2j$:$2j-2$
resonance, and thus can be affected by the second-order in eccentricity
terms.

\subsection{Applications to real and simulated systems}
We next explored how parameter estimates of masses, eccentricities, and longitudes of periastron derived by fitting both real and simulated data with our formulae compared to those found using {\it TTVFast}. We used the second order resonant terms, in combination with the first order eccentric formulae of \citet{Agol2015}, as our analytic model unless otherwise noted. For one simulated system we also used the first order formulae alone for a test. In the following analyses, we employed an affine invariant markov chain monte carlo \citep{GoodmanWeare} to estimate parameters. When fitting real data, we used a student-t likelihood function with 2 degrees of freedom and did not remove outliers. When fitting simulated transit times with Gaussian uncertainties added, we used a Gaussian likelihood function.
{ 
\subsubsection{Kepler-26 (KOI-250)}
The Kepler-26 star hosts four planets with orbital periods of 3.54 (Kepler-26d), 12.28 (Kepler-26b), 17.26 (Kepler-26c), and 46.8 (Kepler-26e) days. The period ratios of adjacent planets are 3.47 (d-b), 1.41 (b-c), and 2.71 (c-e).The innermost planet and outermost planet therefore are not near a low-order mean motion resonance with either middle planet. They do not exhibit TTVs of their own, and we assume that they do not affect the TTVs of the middle two planets either.  In that case, the b-c pair can be treated as an isolated system near the 7:5 resonance. The TTVs of these two planets show periodic behavior on a timescale given by the ``super-period" $2\pi/|7n_2-5n_1|$ as expected, along with a synodic TTV signal \citep{JH_Kep26}.

We modeled the b-c pair using an analytic model that included the 7:5 second order MMR terms and all terms that appear at first and zeroth order in eccentricity. We obtain planet-star mass ratio measurements of $(M_b/M_\oplus) (M_\odot/M_\star) = 9.78 \pm 1.36$ and $(M_c/M_\oplus )(M_\odot/M_\star) = 11.92 \pm 1.38$, which are in close agreement with those obtained in a full dynamical analysis of the data \citep{JH_Kep26}, as shown in Figure \ref{fig:Kep26}. 
\begin{figure}
\centering
\includegraphics[width=\columnwidth]{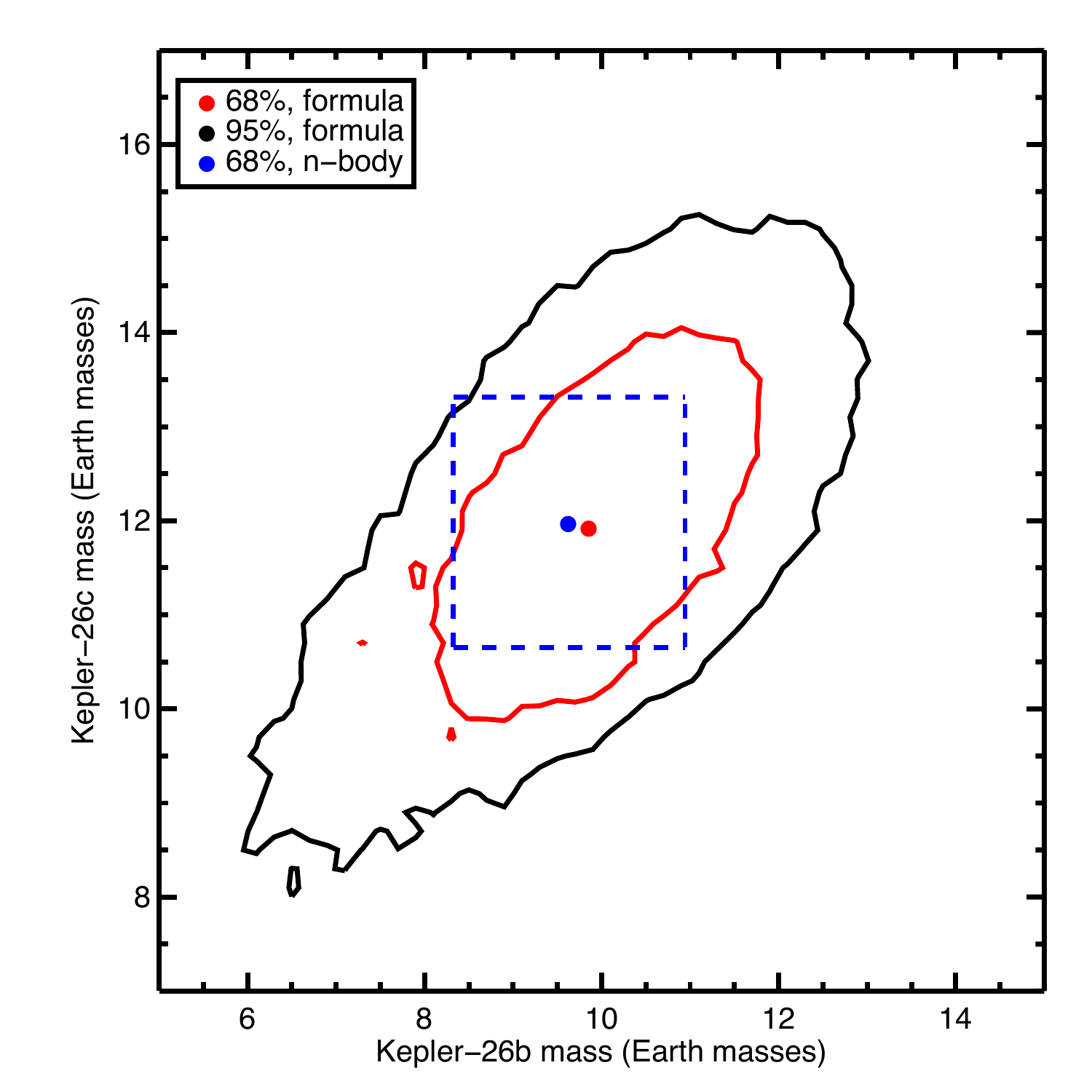}
\caption{Joint posterior probability distribution for the masses of Kepler-26b and Kepler-26c, in Earth masses, assuming a solar mass star. The best fit values from the formula fit are denoted with the red point, with 68\% (red) and 95\% (black) confidence contours shown as well. The blue point and dashed lines reflect the best fit and 68\% boundaries of the \citet{JH_Kep26} fit using a student-t likelihood function.}
\label{fig:Kep26}
\end{figure}

Using the formulae, we find two possible linear correlations between $k_1$ and $k_2$ (where $k_i = e_i \cos{\varpi_i}$) and between $h_1$ and $h_2$ (where $h_i = e_i \sin{\varpi_i}$). These arise since the second order TTV depends approximately on quadratic functions of $\delta k = k_1-k_2$ and $\delta h = h_1-h_2$. That is, $\delta k^2$ and $ \delta h^2$ are singly-peaked, and hence $\delta k$ and $\delta h$ can be either positive or negative, leading to two linear correlations between $k_1$ and $k_2$ (and between $h_1$ and $h_2$). The slope we find which fits these correlations is near unity, as expected based on the arguments of Section \ref{sec:degeneracy}, and the small deviation is related to the error incurred by approximating $g_{j,49}\approx -2 g_{j,45}$ and $g_{j,53} = g_{j,49}$. One of these modes is preferred compared with the other, likely because of (single mode) constraints on $\delta k$ and $\delta h$ resulting from the nearby 4:3 and 3:2 resonances.

\subsubsection{Kepler-46 (KOI-872)}
The star Kepler-46 hosts two transiting planets with orbital periods of 6.8 (Kepler-46d) and 33.6 (Kepler-46b) days. Additionally, Kepler-46b exhibits transit timing variations due to a non-transiting companion (Kepler-46c) near the 5:3 resonance \citep{NesvornyKOI872}. We modeled the transit times of Kepler-46b presented in \citealt{NesvornyKOI872} using the second order terms for the 5:3 resonance, in combination with the full first order formula of \citealt{Agol2015}, and ignoring Kepler-46d. We held the mass of Kepler-46b fixed, as its own mass does not affect its TTVs, but the mass of Kepler-46c and the eccentricities, arguments of pericenter, periods and orbital phases of each planet were allowed to vary. We only searched for a solution near the 5:3 resonance with Kepler-46b, however. 

In Figure \ref{fig:KOI872} we show the results of our formula fit to the transit times of Kepler-46b in comparison with numerical results determined by \citealt{NesvornyKOI872}. We find close agreement in these parameters, as well as in the upper limit of $\sim 0.02$ found for the eccentricity of Kepler-46b (not shown).  Given the best fit period ratio of 1.70, the expected super-period for the 5:3 MMR is nearly 18 orbits of Kepler-46b. The data extend a baseline of 15 transits, with periodicity of $\sim5-6$ orbits of Kepler-46b, and no clear large amplitude signal at a period of 18 orbits. The constraints in this system therefore likely come entirely from a strong ``chopping" TTV, enhanced by contributions from the distant 3:2 MMR and 2:1 MMR (with contributions at periodicities of several orbital periods, as noted by \citealt{NesvornyKOI872}).  The constraints on the eccentricities likely come from the magnitude of the 3:2 and 2:1 MMR contributions as well as the lack of a large signal due to the 5:3 resonance.
\begin{figure}
\centering
\includegraphics[width=\columnwidth]{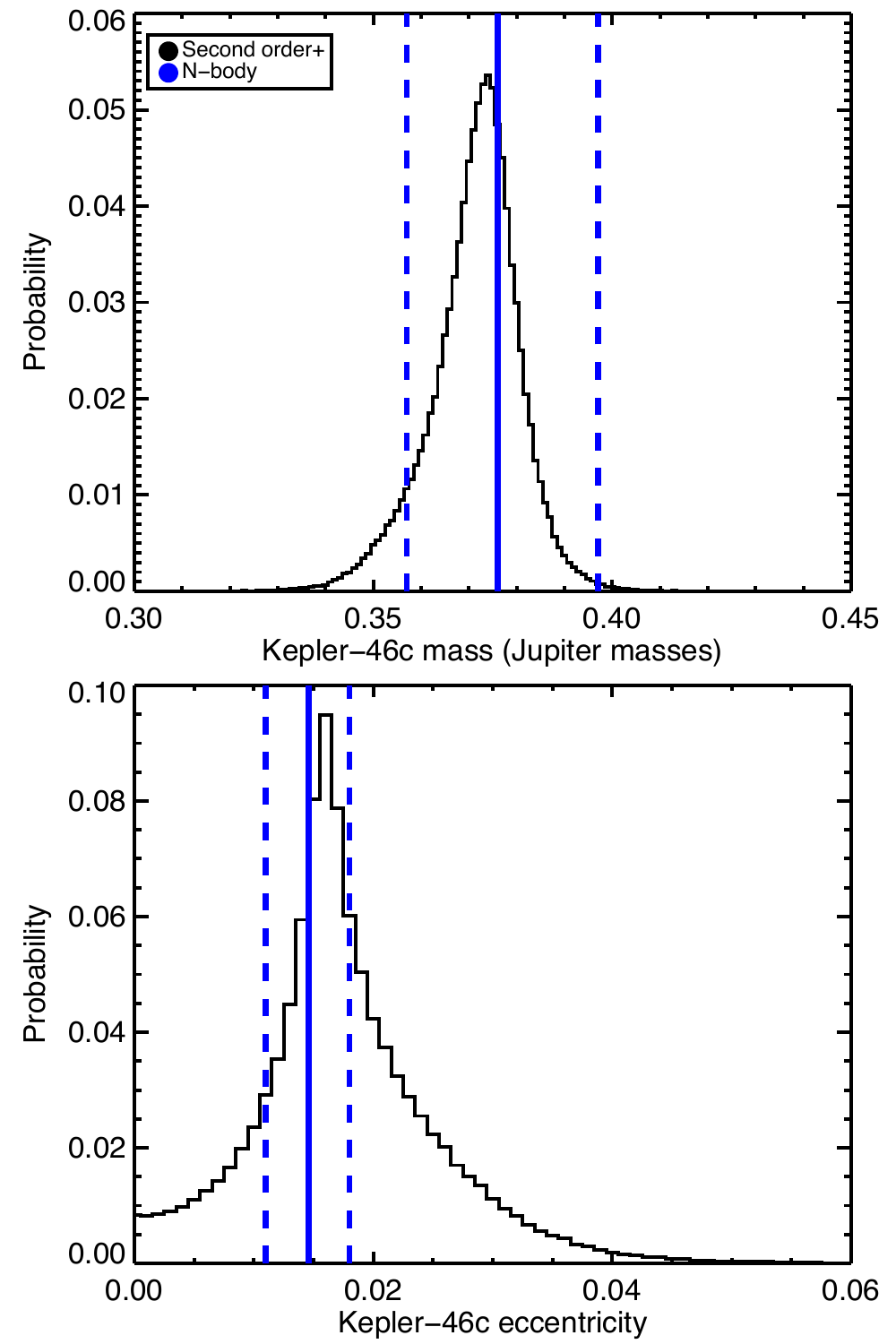}
\caption{Posterior probability distribution for the perturber (Kepler-46c) mass and eccentricity determined using the formula (black). The blue solid line shows the best fit of \citealt{NesvornyKOI872}, and the dashed lines reflect the $\pm 34\%$ confidence contours. }
\label{fig:KOI872}
\end{figure}
}
\subsubsection{Fits to simulated data for a system near a first order MMR, and prospects for measuring individual eccentricities}\label{sec:fit}

Our test case consisted of two 5 Earth mass planets orbiting a solar mass star, with initial osculating periods of 10.0 and 20.2 days, eccentricities of 0.035 and 0.05, and longitudes of pericenter misaligned by $135^\circ$. For these parameters, the system is near the 2:1 resonance, but with important contributions from the 4:2 resonance due to the moderate eccentricities and values of $\delta k$ and $\delta h$. We remark that this system is somewhat similar to KOI-142, which has two planets near a 2:1 resonance with eccentricities of $\sim 0.05$, and which also exhibits TTVs that deviate from a pure sinusoid \citep{KOI142}.

\begin{figure}
\centering
\includegraphics[width=\columnwidth]{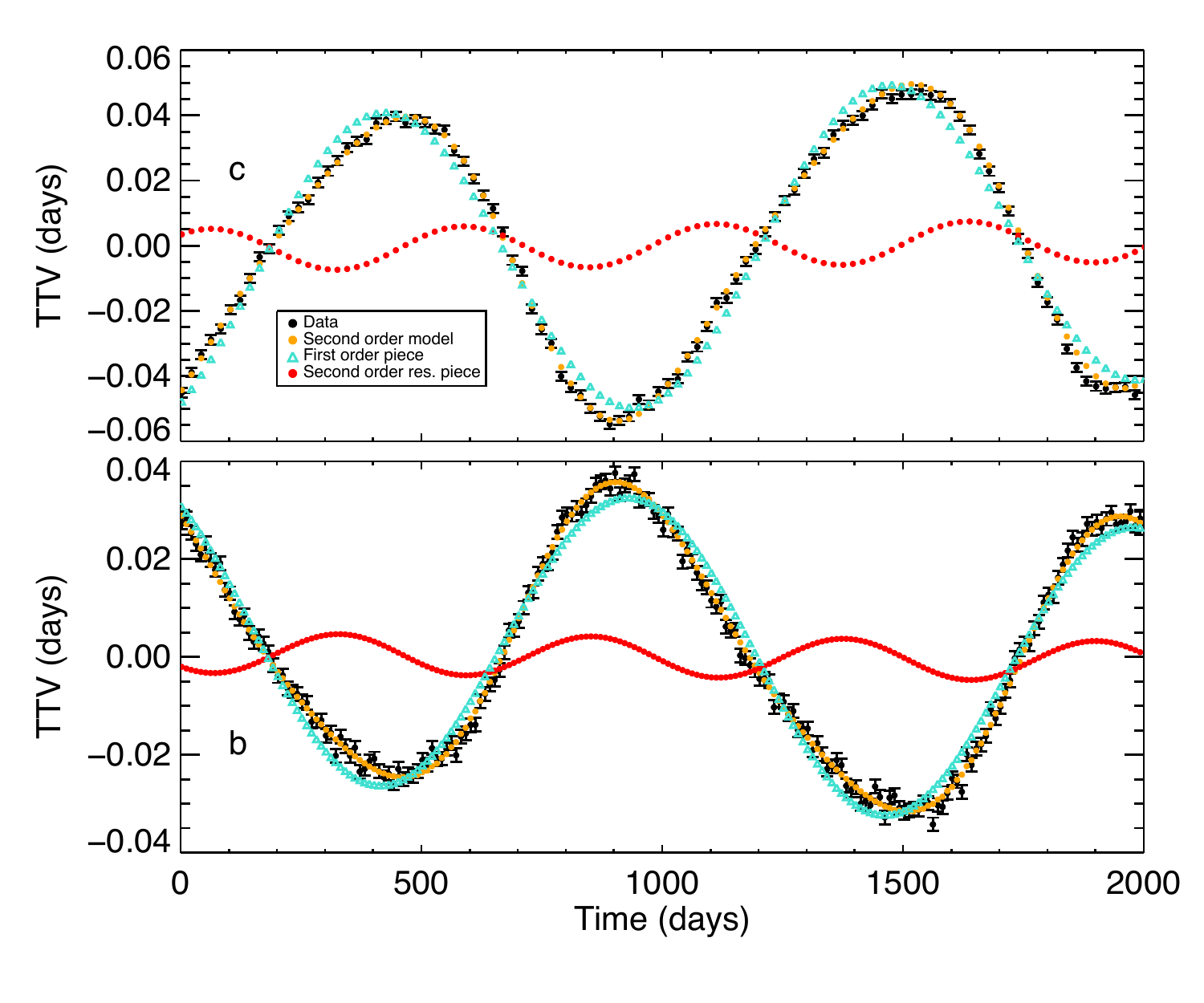}
\caption{Simulated transit timing variations for the higher eccentricity case, in black, a representative solution from the model including the second order and first order terms (from \citealt{Agol2015}), in orange, the contribution to this model from the first order resonant terms alone (turquoise triangles), and the contribution from the second order resonant harmonic (red). Not shown is the individual contribution coming from ``chopping" terms without any small resonant denominators. }
\label{fig:btimes}
\end{figure}

We simulated transit times using {\it TTVFast} and added Gaussian noise with a standard deviation of 2 minutes. In Figure \ref{fig:btimes}, we show the modeled transit timing variations. The various colored points show a sample fit found modeling this data with the second order resonant terms and the first order resonant terms. We also show the contribution of this fit coming from the first order resonant terms alone, as well as from the second order resonant terms alone. (Note that the second order formulae contains the $O(e)$ contribution of the second order resonant terms; we do not double count this.)

\begin{figure}
\centering
\includegraphics[width=\columnwidth]{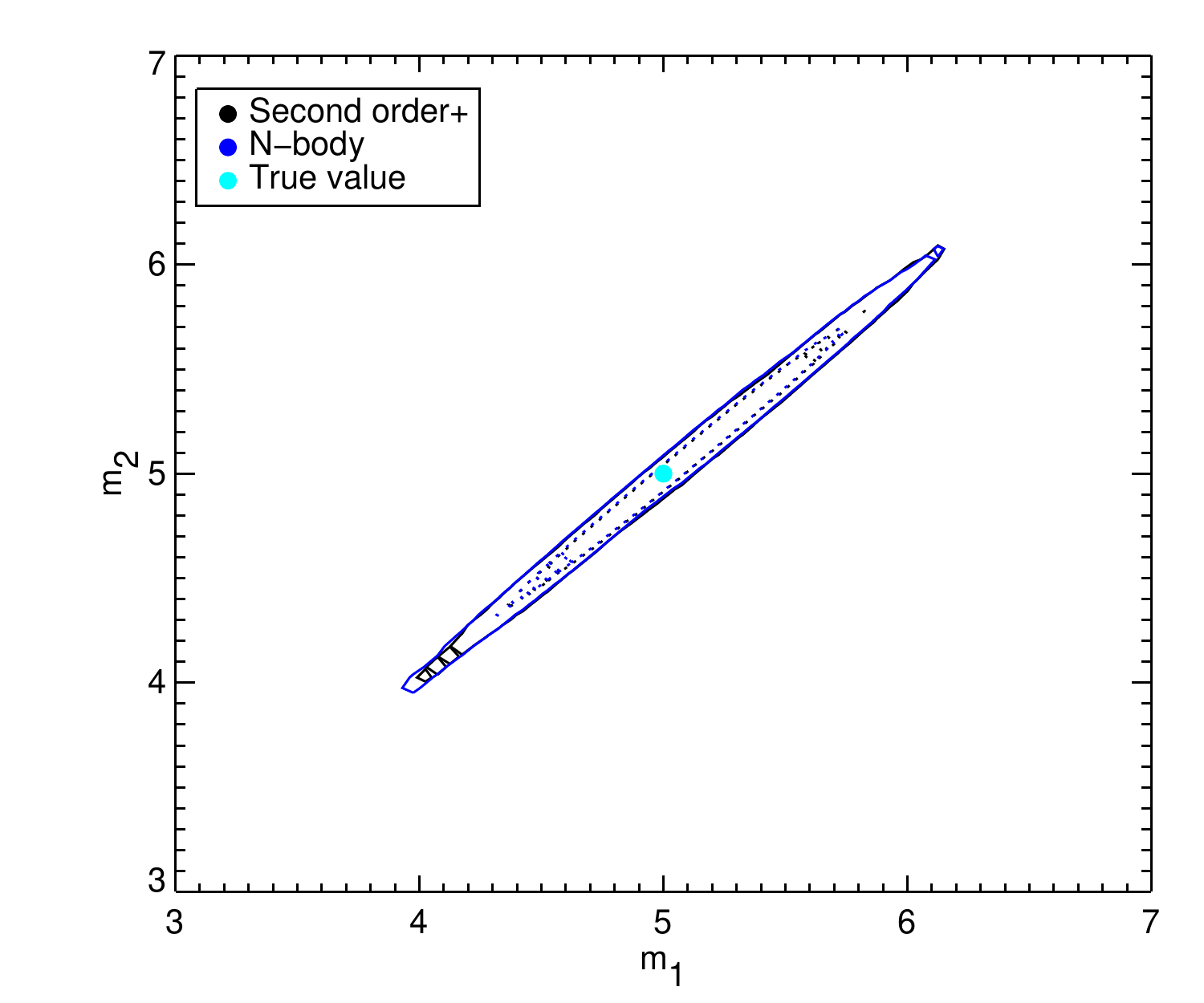}
\caption{Joint posterior 68$\%$ (dotted) and 95$\%$ (solid) confidence contours for the planet masses, in units of Earth mass. The results show are from a full dynamical analysis (blue) and from an analysis using the second order terms in combination with the first order formulae derived in  \citealt{Agol2015} (black).}
\label{fig:bmass}
\end{figure}
\begin{figure}
\centering
\includegraphics[width=\columnwidth]{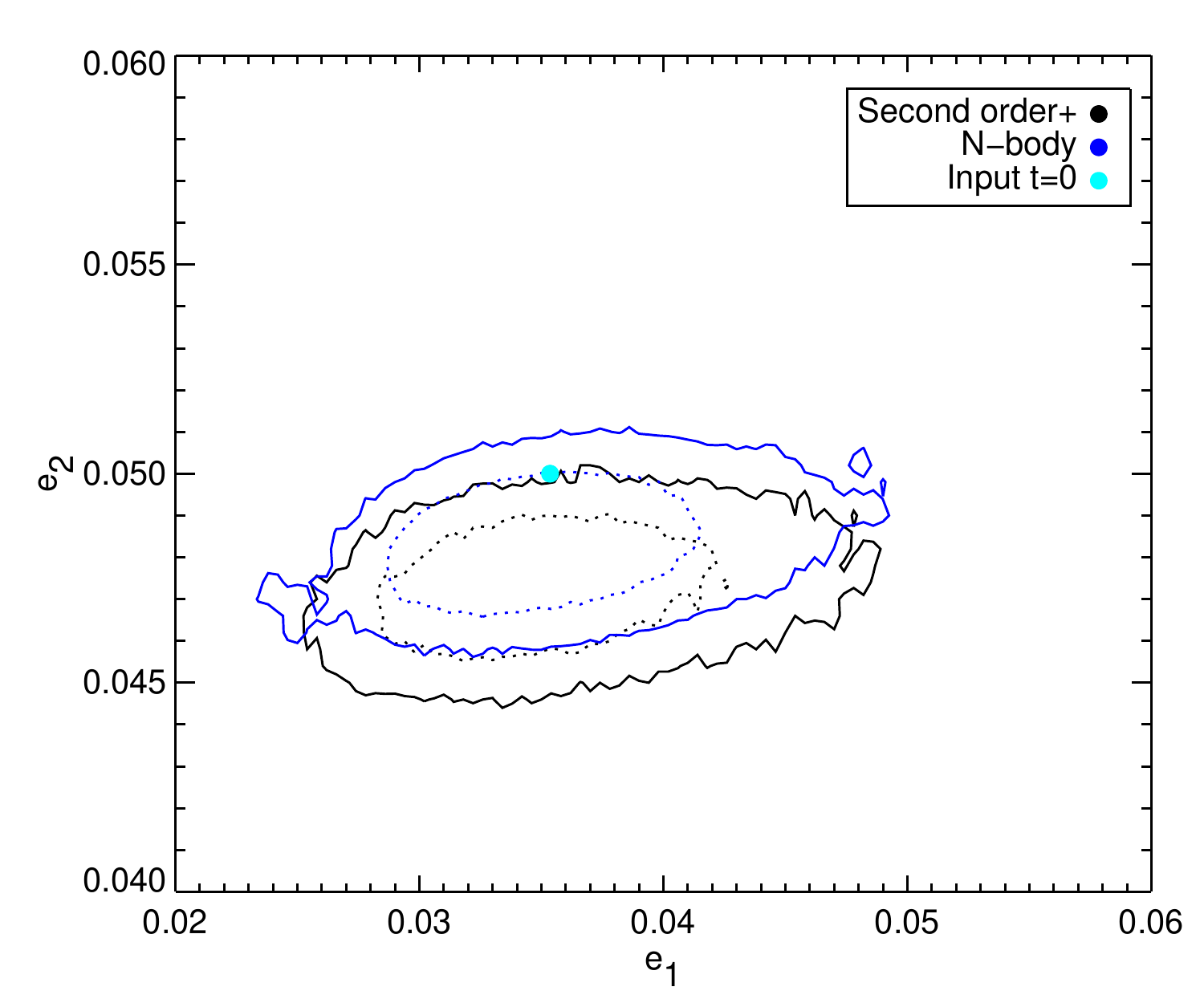}
\caption{Joint posterior 68$\%$ (dotted) and 95$\%$ (solid) confidence contours for the planet eccentricities. The results shown are from a full dynamical analysis (blue) and from an analysis using the second order terms in combination with the first order formulae derived in  \citealt{Agol2015} (black). }
\label{fig:becc}
\end{figure}
In Figure \ref{fig:bmass} and Figure \ref{fig:becc}, we show the resulting joint confidence levels for the two planet masses and eccentricities for both the formula model and for an n-body model. The agreement is very good; we note, however, that modes associated with higher eccentricity were also found using the formula fit, depending on the particular noise realization. However, this multi-modality disappeared as the noise amplitude decreased.  Note also that we used direct n-body integration to simulate the transit times we fit. Hence the input value for the eccentricity is an osculating value. The eccentricities measured via the formula correspond to eccentricities computed from the averaged (canonical) $x$ and $y$ variables, or free eccentricities, while those of the n-body model are initial osculating elements. The difference between the two is on the order of the magnitude of the forced eccentricity due to the near resonance, and may explain why there is a small offset between the numerical and n-body fits.

For these parameters and signal to noise, the eccentricities (and longitudes of pericenter, not shown) are both measured independently. As discussed in Section \ref{sec:degeneracy}, the first order resonant terms as presented in \citet{Lithwick2012} alone suffer from an absolute degeneracy, in that the TTV amplitude and phase depends only on quantities approximately equal to $\delta h = e_1 \cos{\varpi_1}-e_2\cos{\varpi_2}$ and $\delta k = e_1 \sin{\varpi_1}-e_2\sin{\varpi_2}$ (the real and imaginary parts of $Z_{free}$ in the notation of \citealt{Lithwick2012}). Though both the first and second order TTV harmonics depend approximately only on $\delta h$ and $\delta k$, in reality, they both depend on slightly different functions of the eccentricity and pericenter\footnote{This is true especially for the 2:1 MMR, which contains indirect terms in the coefficients of the first order resonant terms, as noted by \citet{Hadden}.}.  Hence at high signal to noise the two harmonics may be used to measure eccentricities and pericenters individually, as in Figure \ref{fig:becc}. Similarly, we hypothesize that the moderate eccentricities of the KOI-142 system produce a detectable second harmonic in the TTVs of KOI-142b, which, in combination with the short period chopping and the transit duration variations, help lead to a unique solution for the non-transiting perturber \citep{KOI142}.

Note that a model neglecting the second harmonic fails to determine the eccentricities correctly: it returns a decent fit, but at significantly higher eccentricities. Including the second order terms in the model leads to the correct answer, but as mentioned there can be multi-modality. It is possible that including the 6:3 resonant harmonic in the fit (an $O(e^3)$ correction, see below) would alleviate this.

\begin{figure}
\centering
\includegraphics[width=\columnwidth]{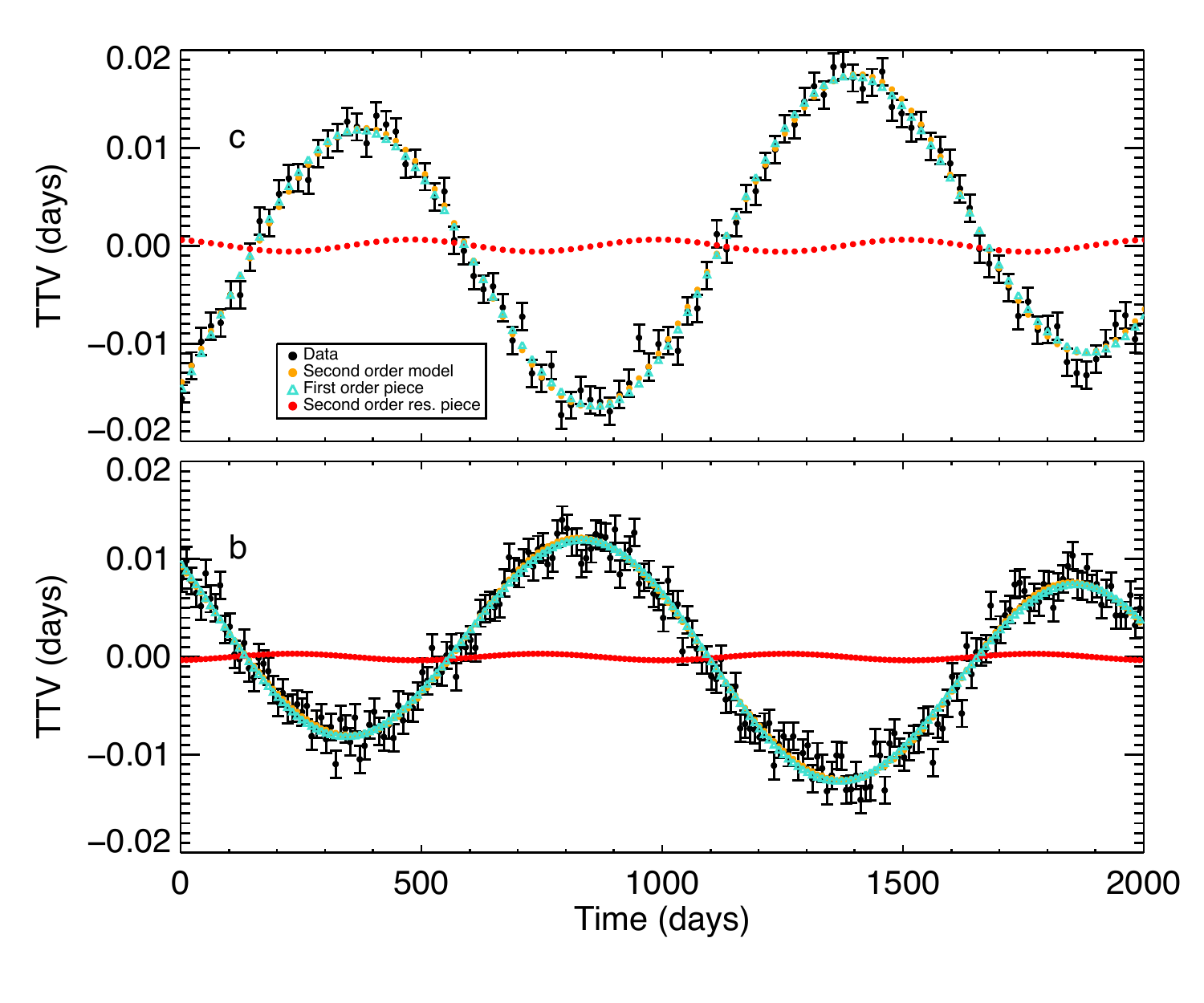}
\caption{Simulated transit timing variations for the lower eccentricity case, in black, a representative solution from the model including the second order and first order terms (from \citealt{Agol2015}), in orange, the contribution to this model from the first order resonant terms alone (turquoise triangles), and the contribution from the second order resonant harmonic (red). }
\label{fig:dtimes}
\end{figure}

To explore how the second order harmonic can lead to an eccentricity measurement, as in Figure \ref{fig:becc}, we decreased the eccentricities to 0.014 and 0.01 in order to reduce the amplitude of the second order harmonic. All other parameters remained the same. In this case, a 2 minute amplitude for the noise is 2-4x larger than the amplitude of the second order harmonic for the two planets. Figure \ref{fig:dtimes} shows the transit times we modeled in addition to a sample fit, again delineating between the entire second order model and the contributions coming from the first order resonant piece and the second order piece.
\begin{figure}
\centering
\includegraphics[width=\columnwidth]{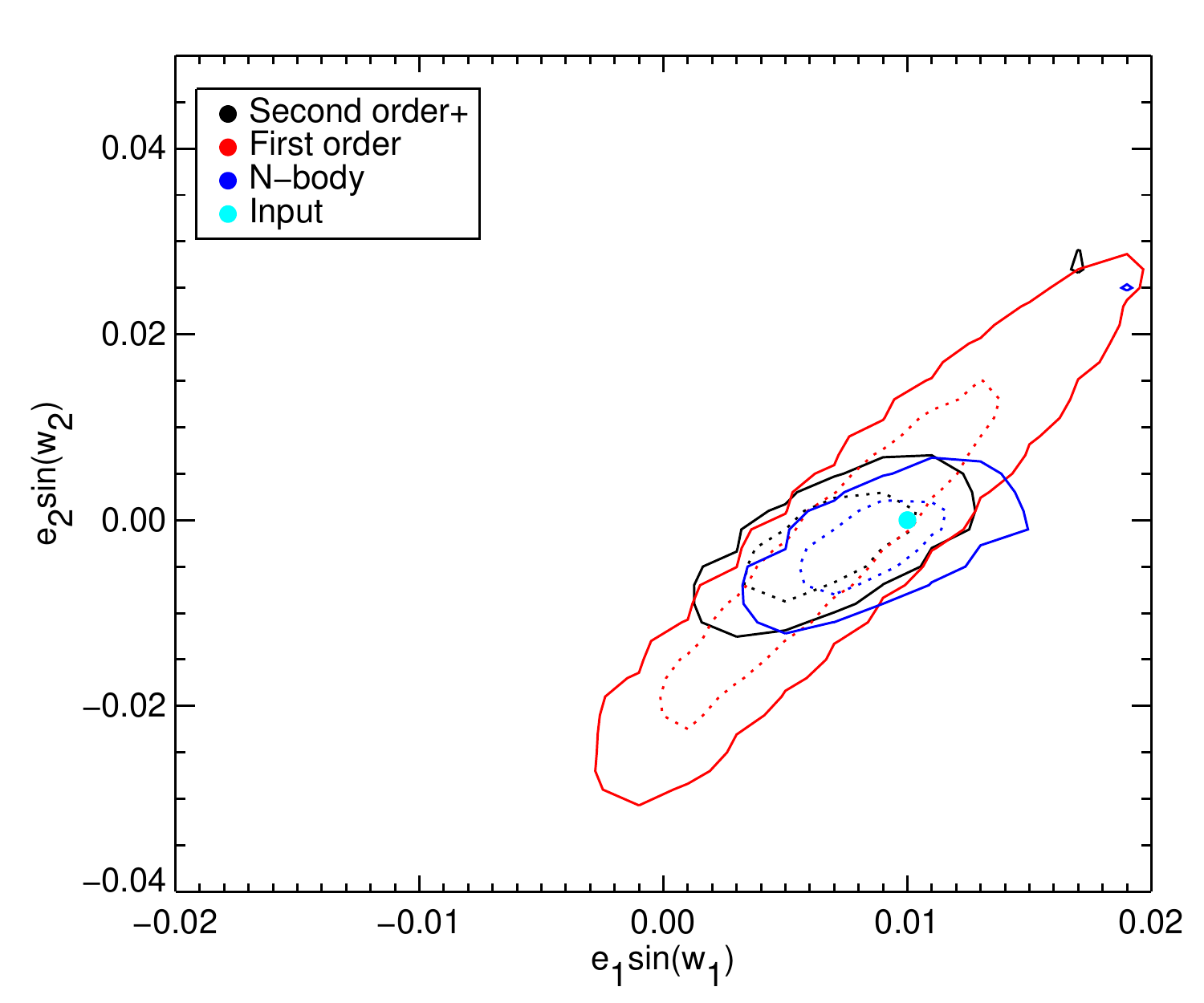}
\caption{Joint posterior 68$\%$ (dotted) and 95$\%$ (solid) confidence contours for the planet eccentricity vector components I. The results shown are from a full dynamical analysis (blue) and from an analysis using the second order terms in combination with the first order formulae derived in  \citealt{Agol2015} (black), and one using only the first order solution (red).}
\label{fig:d_dh_1}
\end{figure}
\begin{figure}
\centering
\includegraphics[width=\columnwidth]{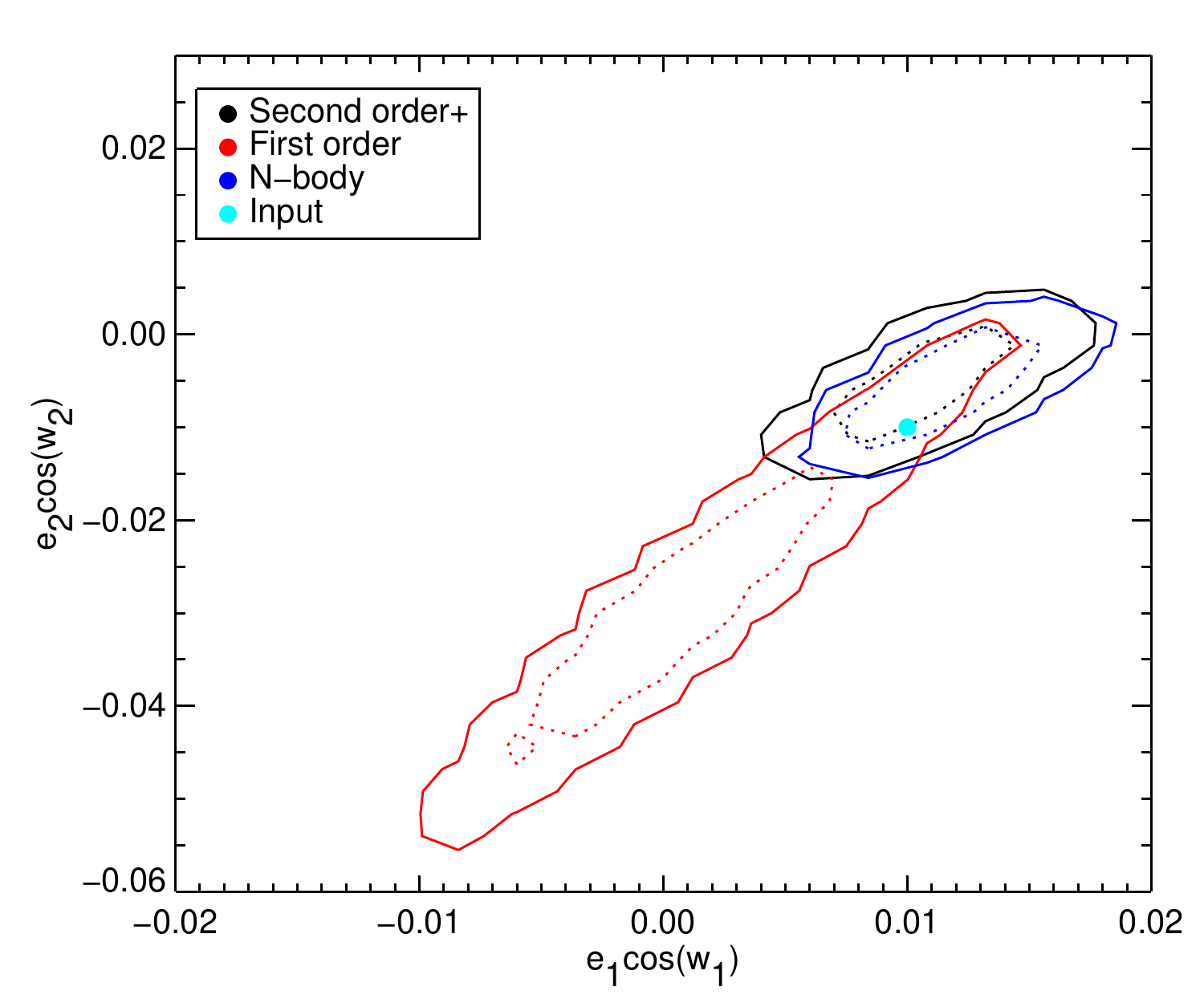}
\caption{Joint posterior 68$\%$ (dotted) and 95$\%$ (solid) confidence contours for the planet eccentricity vector components II. The results shown are from a full dynamical analysis (blue) and from an analysis using the second order terms in combination with the first order formulae derived in  \citealt{Agol2015} (black), and one using only the first order solution (red).}
\label{fig:d_dk_1}
\end{figure}

\begin{figure}
\centering
\includegraphics[width=\columnwidth]{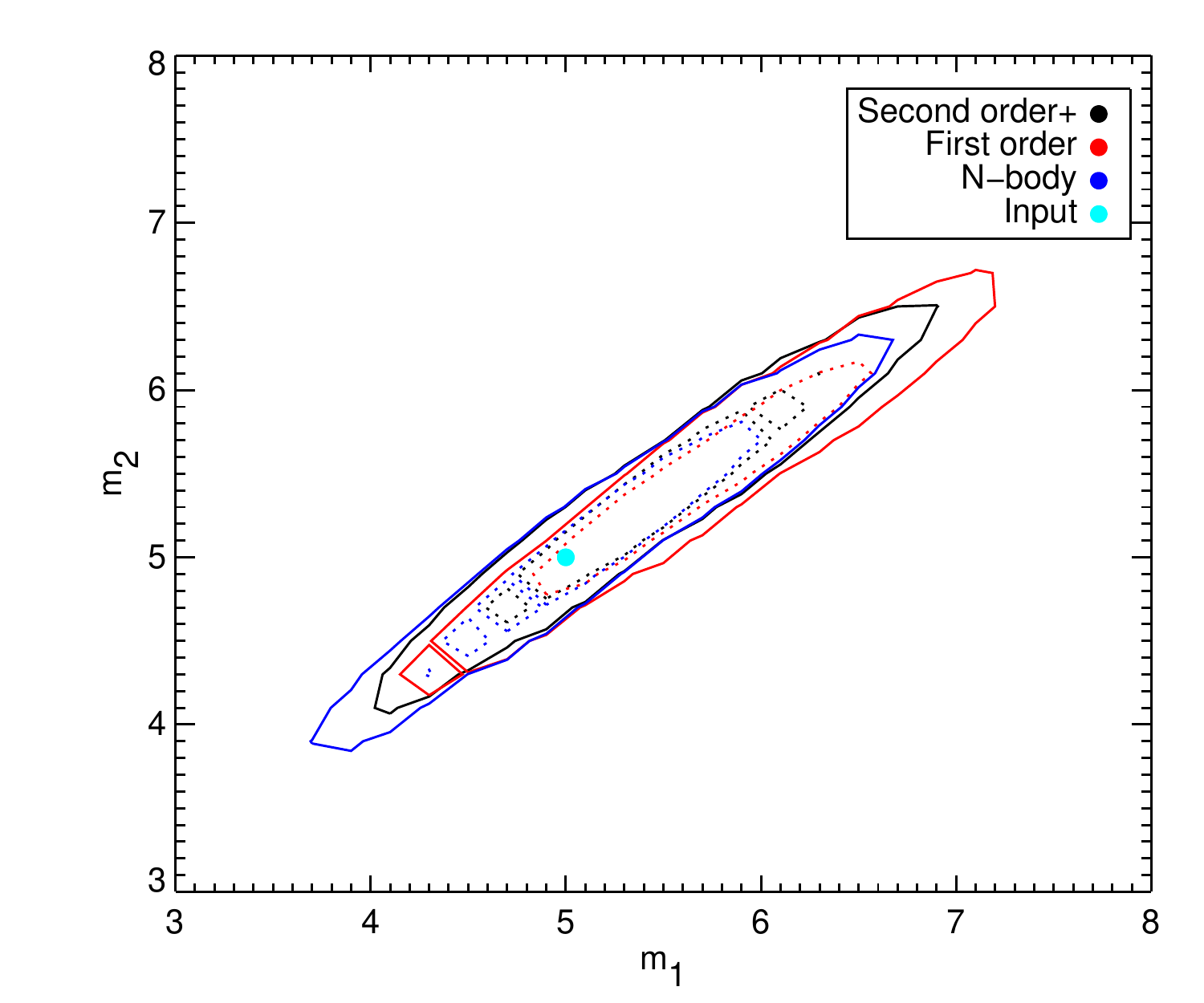}
\caption{ Joint posterior 68$\%$ (dotted) and 95$\%$ (solid) confidence contours for the planet masses, in units of Earth masses. See Figure \ref{fig:d_dh_1} for details.}
\label{fig:mass1}
\end{figure}

Figures \ref{fig:d_dh_1} and \ref{fig:d_dk_1} shows the result of a fit to these times using n-body, using the second and first order eccentricity terms, and now also with only the first order eccentricity formula.   Apparently the second order terms are still a useful constraint on the eccentricities and longitudes of pericenters, even given their low amplitude. This may be because the amplitude of the second order term is constrained to be below the noise. For these particular data, the planet masses are measured equally well by all models, as shown Figure \ref{fig:mass1}.


\section{Higher order resonances}\label{sec:general}
{ We now extend our above derivation to any order eccentricity type resonance. This generalizes the work of \citealt{Boue}, who studied the particular case of a planet on an initially circular orbit and a planet on a fixed eccentric orbit.} In this case, for the $j$:$j-N$ resonance, the Hamiltonian, including {\it only} the resonant terms, takes the form
\begin{align}
H & = -\frac{GM_\star m_1}{2 a_1}-\frac{GM_\star m_2}{2 a_2}-\frac{G m_1 m_2}{a_2}\nonumber \\
&\times \sum_{k=0}^N g_{j,k;N}(\alpha) e_1^ke_2^{N-k} \cos{(\theta_{j;N}-\phi_k)}
\end{align}
where
\begin{align}
\theta_{j;N} & = j\lambda_2-(j-N)\lambda_1,\nonumber \\
\phi_k &= k\varpi_1+(N-k)\varpi_2,
\end{align}
and $g_{j,k;N}(\alpha)$ is a function of Laplace coefficients. For example, for the 2nd order $j$:$j-2$ resonance, $k=0,1,2$, and in that case $g_{j,0;2}= g_{j,53}, g_{j,1;2}= g_{j,49}$ and $g_{j,2;2}= g_{j,45}$. We rewrite the interaction Hamiltonian as
\begin{align}
H_1 & = -\epsilon_1 n_2 \Lambda_2 \sum_{k=0}^N \frac{g_{j,k;N}(\alpha)}{\Lambda_1^{k/2} \Lambda_2^{(N-k)/2}}\bigg(2P_1\bigg)^{k/2} \bigg(2P_2\bigg)^{(N-k)/2} \nonumber \\
&\times \bigg[\Re(e^{i\phi_k})\cos{\theta_{j;N}} +\Im(e^{i\phi_k})\sin{\theta_{j;N}} \bigg]
\end{align}
where $i$ without subscript is $\sqrt{-1}$.

 We can simplify this:
\begin{align}
H_1 & = -\epsilon_1 n_2 \Lambda_2 \sum_{k=0}^N \frac{g_{j,k;N}(\alpha)}{\Lambda_1^{k/2} \Lambda_2^{(N-k)/2}} \nonumber \\
&\times \bigg[ \Re(z_1^k z_2^{N-k})\cos{\theta_{j;N}} -\Im(z_1^k z_2^{N-k})\sin{\theta_{j;N}} \bigg]
\end{align}
where
\begin{align}
z_i & = x_i+i y_i = \sqrt{2P_i}e^{ip_i}
\end{align}
and we have remembered that the canonical angle is not $\varpi$ but $p=-\varpi$. 
This is the exact form of Equation \eqref{Ham2}, with
\begin{align}
\tilde{A}_1 & = \sum_{k=0}^N \frac{g_{j,k;N}(\alpha)}{\Lambda_1^{k/2} \Lambda_2^{(N-k)/2}} \Re(z_1^k z_2^{N-k}) \nonumber \\
\tilde{A}_2 & = -\sum_{k=0}^N \frac{g_{j,k;N}(\alpha)}{\Lambda_1^{k/2} \Lambda_2^{(N-k)/2}} \Im(z_1^k z_2^{N-k})
\end{align}
If we set $N=2$, we find agreement with the expressions for $\tilde{A}_1$ and $\tilde{A}_2$ given in Equation \eqref{Ham2}. We now proceed exactly as above, to find
\begin{align}\label{deviations_general}
\delta \lambda_i &\approx \epsilon_1\frac{n_2 \Lambda_2}{\omega_{j;N}^2}\frac{d\omega_j}{d\Lambda_i}\bigg[\tilde{A}_1 \sin{\theta_{j;N}} - \tilde{A}_2 \cos{\theta_{j;N}}\bigg]  \nonumber \\
 \delta y_i &= - \epsilon_1\frac{n_2 \Lambda_2}{\omega_{j;N}}\bigg[\frac{d\tilde{A}_1}{d x_i} \sin{\theta_{j;N}} - \frac{d\tilde{A}_2}{d x_i} \cos{\theta_{j;N}}\bigg]\nonumber \\
\delta x_i  &=\epsilon_1\frac{n_2 \Lambda_2}{\omega_{j;N}}\bigg[\frac{d\tilde{A}_1}{d y_i} \sin{\theta_{j;N}} - \frac{d\tilde{A}_2}{d y_i} \cos{\theta_{j;N}}\bigg] \nonumber\\
\delta \Lambda_i & \approx 0
\end{align}
with $\omega_{j;N} = jn_2-(j-N)n_1$. 
The deviation in the true longitude is given by \eqref{Eqn3}. We now sketch the derivation for the inner planet:
\begin{align}
&\delta \theta_1  = \epsilon_2 \frac{n_1}{\omega_{j;N}}\alpha\times\bigg[ \sin{\theta_{j;N}} \times\nonumber \\
&\bigg\{\frac{3 n_1 (j-N)}{\omega_{j;N}} \tilde{A}_1+2\sqrt{\Lambda_1}\bigg(\frac{d\tilde{A}_1}{d y_1}\sin{\lambda_1}-\frac{d\tilde{A}_1}{d x_1}\cos{\lambda_1}\bigg)\bigg\} \nonumber \\
& +\cos{\theta_{j;N}}\times \nonumber \\
&\bigg\{-\frac{3 n_1 (j-N)}{\omega_{j;N}} \tilde{A}_2+2\sqrt{\Lambda_1}\bigg(\frac{d\tilde{A}_2}{d x_1}\cos{\lambda_1}-\frac{d\tilde{A}_2}{d y_1}\sin{\lambda_1}\bigg)\bigg\}\bigg] 
\end{align}

The derivative of the real(imaginary) part of a function is the real (imaginary) part of the derivative of the function,  i.e. $\Re(i x) =-\Im(x)$ and $\Im(i x) = \Re(x)$ for a complex number $x$, so we can write:
\begin{align}
\sqrt{\Lambda_1}\frac{d\tilde{A}_1}{d x_1} & = \sum_{k=0}^N g_{j,k;N} k e_1^{k-1} e_2^{N-k}\Re(e^{-i[\phi_k-\varpi_1]}) \nonumber \\
\sqrt{\Lambda_1}\frac{d\tilde{A}_1}{d y_1} & = -\sum_{k=0}^N g_{j,k;N} k e_1^{k-1} e_2^{N-k}\Im( e^{-i[\phi_k-\varpi_1]}) \nonumber \\
\sqrt{\Lambda_1}\frac{d\tilde{A}_2}{d x_1} & = -\sum_{k=0}^N g_{j,k;N} k e_1^{k-1} e_2^{N-k}\Im(e^{-i[\phi_k-\varpi_1]}) \nonumber \\
\sqrt{\Lambda_1}\frac{d\tilde{A}_2}{d y_1} & = -\sum_{k=0}^N g_{j,k;N} k e_1^{k-1} e_2^{N-k}\Re( e^{-i[\phi_k-\varpi_1]})
\end{align}
with analogous expressions for the outer planet (with the pre-factor $k \rightarrow N-k$, $\phi_k-\varpi_1 \rightarrow \phi_k-\varpi_2$, the exponent of $e_1$ becomes $k$, and that of $e_2$ becomes $N-k-1$.).  Written in terms of eccentricities and pericenters, $\tilde{A}_1$ and $\tilde{A}_2$ are

\begin{align}
\tilde{A}_1 & = \sum_{k=0}^N g_{j,k;N}(\alpha) e_1^{k} e_2^{N-k}\Re(e^{-i\phi_k}) \nonumber \\
\tilde{A}_2 & = -\sum_{k=0}^N g_{j,k;N}(\alpha) e_1^{k} e_2^{N-k}\Im(e^{-i\phi_k}),
\end{align}
where $\phi_k = k\varpi_1+(N-k)\varpi_2$.

The final expressions for the deviations in transit times are (after some simplification):
\begin{align}
&\delta t_1  = -\frac{\epsilon_2}{n_1} \frac{n_1}{\omega_{j;N}}\alpha \sum_{k=0}^N g_{j,k;N}(\alpha) e_2^{N-k}e_1^{k-1} \bigg[ \nonumber \\
& \bigg\{\frac{3 n_1 (j-N)}{\omega_{j;N}} e_1 \Re(e^{i\phi_k})-2k \Re(e^{i(\phi_k-\varpi_1+\lambda_1)}) \bigg\}\sin{\theta_{j;N}}\nonumber \\
& +\bigg\{-\frac{3 n_1 (j-N)}{\omega_{j;N}} e_1 \Im(e^{i\phi_k})+2k \Im(e^{i(\phi_k-\varpi_1+\lambda_1)}) \bigg\}\cos{\theta_{j;N}}\bigg] \nonumber \\
&\delta t_2  = -\frac{\epsilon_1}{n_2} \frac{n_2}{\omega_{j;N}} \sum_{k=0}^N g_{j,k;N}(\alpha) e_2^{N-k-1}e_1^{k} \bigg[ \nonumber \\
& \bigg\{-\frac{3 j n_2}{\omega_{j;N}} e_2 \Re(e^{i\phi_k})-2(N-k) \Re(e^{i(\phi_k-\varpi_2+\lambda_2)}) \bigg\}\sin{\theta_{j;N}}+\nonumber \\
& \bigg\{\frac{3 j n_2 }{\omega_{j;N}} e_2 \Im(e^{i\phi_k})+2(N-k) \Im(e^{i(\phi_k-\varpi_2+\lambda_2)}) \bigg\}\cos{\theta_{j;N}}\bigg] 
\end{align}

or, casting in the same symbols as we did for the second order resonances (Equation \eqref{ttvs_a_b}),
\begin{align}\label{ttvs_a_b_general}
\delta t_1&  = -\frac{1}{n_1}\epsilon_2 \alpha\bigg[\bigg\{3(j-N)\bigg(\frac{n_1}{\omega_{j;N}}\bigg)^2 A_1-2\bigg(\frac{n_1}{\omega_{j;N}}\bigg)B^1_1\bigg\}\sin{\theta_{j;N}}\nonumber \\
&+\bigg\{3(j-N)\bigg(\frac{n_1}{\omega_{j;N}}\bigg)^2 A_2-2\bigg(\frac{n_1}{\omega_{j;N}}\bigg)B^1_2\bigg\}\cos{\theta_{j;N}}\bigg]
\end{align}
and
\begin{align}
\delta t_2  &= -\frac{1}{n_2}\epsilon_1 \bigg[\bigg\{-3j\bigg(\frac{n_2}{\omega_{j;N}}\bigg)^2 A_1-2\bigg(\frac{n_2}{\omega_{j;N}}\bigg)B^2_1\bigg\}\sin{\theta_{j;N}}\nonumber \\
&+\bigg\{-3j\bigg(\frac{n_2}{\omega_{j;N}}\bigg)^2 A_2-2\bigg(\frac{n_2}{\omega_{j;N}}\bigg)B^2_2\bigg\}\cos{\theta_{j;N}}\bigg]
\end{align}
where $\theta_{j;N}=j\lambda_2-(j-N)\lambda_1$ and 
\begin{align}
A_1 & = \sum_{k=0}^N g_{j,k;N}(\alpha) e_2^{N-k}e_1^{k}\cos{\phi_k} \nonumber \\
A_2 & = -\sum_{k=0}^N g_{j,k;N}(\alpha) e_2^{N-k}e_1^{k}\sin{\phi_k}\nonumber \\
B_1^1 & = \sum_{k=0}^N g_{j,k;N}(\alpha) ke_2^{N-k}e_1^{k-1}\cos{(\phi_k-\varpi_1+\lambda_1)}\nonumber \\
B_2^1 & = -\sum_{k=0}^N g_{j,k;N}(\alpha) ke_2^{N-k}e_1^{k-1}\sin{(\phi_k-\varpi_1+\lambda_1)}\nonumber \\
B_1^2 & = \sum_{k=0}^N g_{j,k;N}(\alpha) (N-k)e_2^{N-k-1}e_1^{k}\cos{(\phi_k-\varpi_2+\lambda_2)}\nonumber \\
B_2^2 & = -\sum_{k=0}^N g_{j,k;N}(\alpha) (N-k)e_2^{N-k-1}e_1^{k}\sin{(\phi_k-\varpi_2+\lambda_2)}
\end{align}
with $\phi_k = k\varpi_1+(N-k)\varpi_2$. 

Therefore, close enough to any eccentricity type resonance - with the caveat that the system is not in the resonance and that the higher order eccentricity terms neglected are small - the TTVs  of a pair of planets are periodic with a timescale of $2\pi/|jn_2-(j-N)n_1|$. The amplitudes depend linearly on the mass of the perturbing planet, relative to the mass of the star, and on the eccentricities and pericenters, as well as on the mean longitude of the transiting planet at transit. The phases also depend on these quantities, though they are independent of the masses. However, the amplitude and phase of these TTVs do not uniquely constrain the masses, eccentricities, and pericenters, since this amounts to six parameters and only four observables. 

 We hypothesize that the TTVs will, in the limit of compact orbits (higher $j$ for a given $N$), be anti-correlated, only depend on approximately $\delta k$ and $\delta h$, and that the $N-$th order resonant TTV will include powers up to $|\vec{e}|^N$, with $\vec{e} = (\delta k, \delta h)$. Physically, this dependence makes sense since anti-aligned orbits allow for closer approaches and stronger interactions between the planets. Mathematically this maximizes $|\vec{e}|$ to produce larger TTVs.

In the above derivation, we neglected terms of order $e(e^{N-1}/\delta)$ (where again $\delta$ is a normalized distance to resonance, defined in Equation \eqref{small_params}, and we assume it is small). More importantly, neglected chopping effects, without any small denominators, appear at every order in $e$. If one combines the $N-$th order resonant TTV with the $e^1$ chopping TTV formulae \citep{Agol2015}, one still will find errors at low eccentricity. This arises because $e$ must be larger for higher order resonances to be important, and in this case neglected chopping terms at much lower powers of $e$ may also be important. Hence the $N-$th order TTV formulae above may be of limited use, even if combined with other known formulae. 

As an exercise, we can use these formulae to confirm those of \citet{Lithwick2012}. In that case, $N=1$, and 
\begin{align}
A_1 & =  g e_2\cos{\varpi_2}+f e_1\cos{\varpi_1} = \Re(Z_{free}^\star) \nonumber \\
A_2 & =-g e_2\sin{\varpi_2}-f e_1\sin{\varpi_1} =  \Im(Z_{free}^\star)  \nonumber \\
B_1^1 & = f\nonumber \\
B_2^1 & = 0\nonumber \\
B_1^2 & =  g\nonumber \\
B_2^2 & = 0
\end{align}
setting $Z_{free} = fe_1e^{i\varpi_1}+ge_2e^{i\varpi_2}$, $f = g_{j,k=1;N}$ and $g=g_{j,k=0;N}$ and approximating $\lambda$(transit)=0. The TTVs then are
\begin{align}\label{ttvs_a_b_first}
\delta t_1&  = -\frac{2}{n_1}\epsilon_2 \alpha^{-1/2}\frac{1}{j\Delta}\bigg[\bigg\{\frac{3(j-1)}{2j \Delta}\alpha^{-3/2} \Re(Z_{free}^\star)+f\bigg\}\sin{\theta_j}\nonumber \\
&+\bigg\{\frac{3(j-1)}{2j \Delta}\alpha^{-3/2}   \Im(Z_{free}^\star)\bigg\}\cos{\theta_{j}}\bigg]
\end{align}
and
\begin{align}\label{ttvs_a_b_first2}
\delta t_2  &= \frac{2}{n_2}\epsilon_1 \bigg(\frac{1}{j\Delta}\bigg)\bigg[\bigg\{\frac{3}{2\Delta} \Re(Z_{free}^\star)-g\bigg\}\sin{\theta_{j}}\nonumber \\
&+\bigg\{\frac{3}{2\Delta} \Im(Z_{free}^\star)\bigg\}\cos{\theta_{j}}\bigg]
\end{align}
with $\Delta = (j-1)/j(P_2/P_1)-1$.

In the \citet{Lithwick2012} paper, the TTVs are written in the form
\begin{align}
\delta t  &= \frac{V}{2i}e^{i\theta_j}+c.c \nonumber \\
& = \Re(V)\sin{\theta_j}+\Im(V)\cos{\theta_j}
\end{align}
Equations \eqref{ttvs_a_b_first}  and \eqref{ttvs_a_b_first2} take that form if
\begin{align}
V_1 & =-\frac{P_1}{\pi}\frac{\epsilon_2}{j\Delta} \alpha^{-1/2} \bigg(f+ \frac{3}{2\Delta}\frac{(j-1)}{j\alpha^{3/2}} Z_{free}^\star\bigg)\nonumber \\
V_2 & = \frac{P_2}{\pi}\frac{\epsilon_1}{j\Delta}\bigg(-g+\frac{3}{2\Delta}Z_{free}^\star \bigg)
\end{align}
which are equivalent to (A.28) and (A.29) of \citet{Lithwick2012}.

\section{Conclusion}\label{sec:conclude}
We have derived an expression for the TTVs of a pair of planets near the $j$:$j-2$ second order mean motion resonance in the regime of low eccentricities. In this case, the TTV of each planet is sinsuoidal, with a frequency of $jn_2-(j-2)n_1$,  an amplitude linearly dependent on the mass of the perturbing planet, relative to the mass of the star, and with both amplitude and phase dependent on a function of the eccentricities and longitudes of pericenter. In this case, there are six parameters but only four observables, yielding (in principle) mass measurements but not unique eccentricity and pericenter measurements. We show that the same is true for higher order eccentricity-type resonances. This second result, however, may not be of (much) practical use since few pairs are found very near high order mean motion resonances where the formulae apply. However, it does illustrate that TTVs of systems near an $N-$th order resonance will appear with a period given by the super period $2\pi/|jn_2-(j-N)n_1|$ and therefore that higher order eccentricity corrections to the TTVs of planets near first order resonances appear at harmonics of the fundamental (super) period $= 1/|j/P_2-(j-1)/P_1|$.

 At a further level of approximation, which will be relevant for low signal-to-noise data, we have shown that the TTVs of two planets near a second order resonance are anti-correlated. In this case, there is an explicit degeneracy between masses and the combinations $e_1\cos{\varpi_1}-e_2\cos{\varpi_2}$ and $e_1\sin{\varpi_1}-e_2\sin{\varpi_2}$. This result is entirely analogous to that found for first order resonances. We hypothesize that this basic result extends to higher order eccentricity-type resonances.
 
 In order to alleviate the degeneracies between parameters that result for near resonant systems, one must measure a different component of the TTV. This could be the chopping signal, associated with each planet conjunction, which primarily depends on the masses of the planets.  The higher order correction associated with the $2k$:$2k-2$ second order resonance derived here can help constrain eccentricities of planets near the $k$:$k-1$ first order resonance, though high signal-to-noise data is likely required to lead to precise individual eccentricity measurements. When modeling TTVs, we find it helpful to consider the number of significantly measured observables in a TTV (in terms of amplitudes, phases, etc. of different harmonics) in comparison with the number of free parameters, in light of intrinsic degeneracies which TTV formulae help to illuminate.

\acknowledgements
We would like to thank our referee who helped us to clarify and improve this document. KD acknowledges support from the JCPA postdoctoral fellowship at Caltech. EA acknowledges support from NASA grants NNX13AF20G, NNX13AF62G, and NASA Astrobiology Institutes Virtual Planetary Laboratory, supported by NASA under cooperative agreement NNH05ZDA001C. 
\bibliographystyle{apj}
\bibliography{OE2}

\end{document}